\documentclass[11pt]{article}
\usepackage{amsmath,amssymb}
\usepackage{natbib,graphicx,psfrag,epsfig,subfigure,dsfont,times}
\usepackage{authblk}
\usepackage{hyperref,float,helvet}
\usepackage[small]{caption}
\usepackage{color,multicol}
\usepackage{natbib}
\usepackage{multirow}
\usepackage{color, colortbl}
\definecolor{Gray}{gray}{0.9}
\usepackage{setspace}
\graphicspath{{Figures/}}

\newcommand{\dd}{\mathrm{d}}

\newcommand{\lb}{\left[}
\newcommand{\rb}{\right]}
\newcommand{\lc}{\left\{}
\newcommand{\rc}{\right\}}
\newcommand{\be}{\begin{eqnarray}}
\newcommand{\ee}{\end{eqnarray}}

\newcommand{\bm}[1]{\boldsymbol{#1}}

\begin{document}

\title{A New Family of Error Distributions for Bayesian Quantile Regression}
\author{Yifei Yan and Athanasios Kottas
\thanks{
Y. Yan (yifeiyan@soe.ucsc.edu) is Ph.D. student, and A. Kottas (thanos@soe.ucsc.edu)
is Professor of Statistics, \\
Department of Applied Mathematics and Statistics, University
of California, Santa Cruz, CA, 95064, USA.}
}

\date{}

\maketitle

\begin{abstract}
\noindent
We propose a new family of error distributions for model-based quantile
regression, which is constructed through a structured mixture of normal
distributions. The construction enables fixing specific percentiles of
the distribution while, at the same time, allowing for varying mode,
skewness and tail behavior. It thus overcomes the severe limitation of
the asymmetric Laplace distribution -- the most commonly used error
model for parametric quantile regression -- for which the skewness of
the error density is fully specified when a particular percentile is fixed.
We develop a Bayesian formulation for the proposed quantile regression
model, including conditional lasso regularized quantile regression
based on a hierarchical Laplace prior for the regression coefficients,
and a Tobit quantile regression model.
Posterior inference is implemented via Markov Chain Monte Carlo methods.
The flexibility of the new model relative to the asymmetric Laplace
distribution is studied through relevant model properties, and 
through a simulation experiment to compare the two error distributions in 
regularized quantile regression. Moreover, model 
performance in linear quantile regression, regularized quantile
regression, and Tobit quantile regression is illustrated with data
examples that have been previously considered in the literature.
\end{abstract}

regression; Skew normal distribution; Tobit quantile regression.

\newpage

\section{Introduction}

Quantile regression offers a practically important alternative to traditional
mean regression, and forms an area with a rapidly increasing literature.
Parametric quantile regression models are almost exclusively built from the
asymmetric Laplace (AL) distribution the density of which is
\begin{eqnarray}
\label{AL-pdf}
f^{\text{AL}}_{p}(y \mid \mu,\sigma) & = & \frac{p(1-p)}{\sigma}
\exp \left\{- \frac{1}{\sigma} \rho_{p} \left( y - \mu \right)\right\},
\,\,\,\,\, y \in \mathbb{R}
\end{eqnarray}
where $\rho_{p}(u)=$ $u[p-I(u<0)]$, with $I(\cdot)$ denoting the indicator
function. Here, $\sigma >0$ is a scale parameter, $p \in (0,1)$,
and $\mu \in \mathbb{R}$ corresponds to the $p$th percentile,
$\int_{-\infty}^{\mu} f^{\text{AL}}_{p}(y \mid \mu,\sigma) \text{d}y=p$.
Hence, a model for $p$th quantile regression can be developed by expressing
$\mu$ as a function of available covariates $\bm{x}$, for instance, $\mu=$
$\bm{x}^{T} \bm{\beta}$ yields a linear quantile regression structure. Note that
maximizing the likelihood with respect to $\bm{\beta}$ under an AL
response distribution corresponds to minimizing for $\bm{\beta}$ the
check loss function, $\sum_{i=1}^{n} \rho_{p}(y_{i} - \bm{x}^{T}_{i} \bm{\beta})$,
used for classical semiparametric estimation in linear quantile regression
\citep{Koenker2005}.

The AL distribution is receiving increasing attention
in the Bayesian literature, originating from work on inference
for linear quantile regression \citep{YuMoye2001,Tsionas2003}.
Particularly relevant to the Bayesian framework are the different
mixture representations of the distribution \citep{KotzKozu2001},
which have been exploited to construct posterior simulation algorithms
\citep{KozuKoba2011}, as well as to explore different modeling
scenarios; see, for instance, \cite{LG2012} and \cite{WKYLF2013}.

However, the AL distribution has substantial limitations
as an error model for quantile regression. Most striking is that the skewness
of the error density is fully determined when a specific percentile is
chosen, that is, when $p$ is fixed. In particular, the error density is
symmetric in the case of median regression, since for $p=0.5$, the AL
reduces to the Laplace distribution. Moreover,
the mode of the error distribution is at zero, for any $p$, which
results in rigid error density tails for extreme percentiles.
%
%

The literature includes Bayesian nonparametric models for the
error distribution in the special case of median regression
\citep{WM2009,KG2001,HJ2002} and in general quantile regression
\citep{AKMK2009,RBW2010}. The Bayes nonparametrics literature
has also explored inference methods for simultaneous quantile
regression \citep{TK2010,TokdarKadane2012,RS2013}.
However, work on parametric alternatives to AL quantile regression
errors is limited, and the existing models do not overcome all the
limitations discussed above. For instance, although the class of skew
distributions studied in \cite{WCG2014} includes the AL as a special case, it
shares the same restriction with the AL as a quantile regression error
model in that it has a single parameter that controls both skewness
and percentiles. 
\cite{ZhuZind2009} and \cite{ZhuGalb2011} explored the family of 
asymmetric exponential power distributions, which does not include 
the AL distribution. For a fixed probability $p$, the density
function has four free parameters and allows for different decay rates
in the left and the right tails. However, similar to the AL, the mode
of the distribution is fixed at the quantile $\mu$ by construction.

More flexible parametric quantile regression error models
are arguably useful both to expand the inferential scope of the
asymmetric Laplace in the standard quantile regression setting, as
well as to provide building blocks for model development under
more complex data structures. The limited scope of results in this direction
may be attributed to the challenge of defining sufficiently flexible
distributions that are parameterized by percentiles and, at the same time,
allow for practicable modeling and inference methods.

Seeking to fill this gap, we propose a new family of distributions
that is parameterized in terms of percentiles, and overcomes the
restrictive aspects of the AL distribution. The distribution
is developed constructively through an extension of an AL 
mixture representation. In particular, we
introduce a shape parameter to obtain a distribution that has more
flexible skewness and tail behaviour than the AL
distribution, while retaining it as a special case of the new model.
The latter enables connections with the check loss function which are
useful in studying the utility of the new model in the context of
regularized quantile regression. 
Owing to its hierarchical mixture representation, the proposed distribution
preserves the important feature of ready to implement
posterior inference for Bayesian quantile regression.

In Section 2, we develop the new distribution and discuss its
properties relative to the AL distribution. In Section 3, we formulate
the Bayesian quantile regression model, including a prior
specification for the regression coefficients that encourages
shrinkage resulting in regularized quantile regression, and a Tobit
quantile regression formulation. In Section 4, we present results from
a simulation study to compare the performance of the AL and the proposed 
distribution in regularized quantile regression. The methodology is 
illustrated with three data examples in Section 5, focusing again on
comparison with the AL quantile regression model.
Finally, Section 6 concludes with a summary and discussion
of possible extensions.

\section{The generalized asymmetric Laplace distribution}

%
%

The construction of the new distribution is motivated by the most
commonly used mixture representation of the AL density. In particular,
\begin{equation}
\label{AL-mixture}
f^{\text{AL}}_{p}(y \mid \mu,\sigma)  =  \int_{\mathbb{R}^{+}}
\text{N}(y \mid \mu + \sigma A(p) z, \sigma^{2} B(p) z)
\, \text{Exp}(z \mid 1) \, \dd z
\end{equation}
where $A(p) = (1-2p)/\{ p(1-p) \}$ and $B(p) = 2/\{ p(1-p) \}$.
Moreover, $\text{N}(m,W)$ denotes the normal distribution with mean $m$ and
variance $W$, and $\text{Exp}(1)$ denotes the exponential distribution
with mean $1$. We use such notation throughout to indicate either the
distribution or its density, depending on the context.

The mixture formulation in (\ref{AL-mixture}) enables exploration of
extensions to the AL distribution. Extending the $\text{Exp}(1)$
mixing distribution is not a fruitful direction in terms of evaluation
of the intergal, and, more importantly, with respect to fixing percentiles of
the resulting distribution. However, both goals are accomplished by
replacing the normal kernel in (\ref{AL-mixture}) with a skew normal
kernel \citep{Azza1985}.
In its original parameterization, the skew normal density is given by
$f^{\text{SN}}(y \mid \xi,\omega,\lambda)=$
$2 \omega^{-1} \, \phi( \omega^{-1} (y-\xi)) \, \Phi( \lambda \omega^{-1} (y-\xi))$,
where $\phi(\cdot)$ and $\Phi(\cdot)$ denote the density and distribution
function, respectively, of the standard normal distribution. Here, $\xi \in \mathbb{R}$
is a location parameter, $\omega > 0$ a scale parameter, and
$\lambda \in \mathbb{R}$ the skewness parameter. Key to our
construction is the fact that the skew normal density can be written
as a location normal mixture with mixing distribution given by a
standard normal truncated on $\mathbb{R}^{+}$ \citep{Henz1986}.
More specifically, reparameterize $(\xi,\omega,\lambda)$ to
$(\xi,\tau,\psi)$, where $\tau > 0$ and $\psi \in \mathbb{R}$, such
that $\lambda=$ $\psi/\tau$ and $\omega=$ $(\tau^{2} + \psi^{2})^{1/2}$.
Then, $f^{\text{SN}}(y \mid \xi,\tau,\psi)=$
$\int_{\mathbb{R}^{+}} \text{N}(y \mid \xi + \psi s,\tau^{2}) \text{N}^{+}(s \mid 0,1)
\, \dd s$, where $\text{N}^{+}(0,1)$ denotes the standard normal
distribution truncated over $\mathbb{R}^{+}$.

The proposed model, referred to as generalized asymmetric Laplace (GAL)
distribution, is built by adding a shape parameter, $\alpha \in \mathbb{R}$,
to the mean of the normal kernel in (\ref{AL-mixture}) and mixing with respect to
a $\text{N}^{+}(0,1)$ variable. More specifically,
the full mixture representation for the density function,
$f(y \mid p,\alpha,\mu,\sigma)$, of the new distribution is as follows
\begin{eqnarray}
\label{GAL-mixture}
\iint_{\mathbb{R}^{+} \times \mathbb{R}^{+}}
\text{N}(y \mid \mu + \sigma \alpha s + \sigma A(p) z, \sigma^{2} B(p) z)
\, \text{Exp}(z \mid 1) \, \text{N}^{+}(s \mid 0,1) \, \dd z \dd s.
\end{eqnarray}
Note that, integrating over $s$ in (\ref{GAL-mixture}), the GAL
density can be expressed in the form of
(\ref{AL-mixture}) with the $\text{N}(y \mid \mu + \sigma A(p) z, \sigma^{2} B(p) z)$
kernel replaced with a skew normal kernel, which, in its original
parameterization, has location parameter $\mu + \sigma A(p) z$, scale
parameter $\sigma \{ \alpha^{2} + B(p) z \}^{1/2}$, and skewness parameter
$\alpha \{ B(p) z \}^{-1/2}$. Evidently, when $\alpha=0$, $f(y \mid p,0,\mu,\sigma)$
reduces to the AL density.

To obtain the GAL density,
we integrate out first $z$ and then $s$ in (\ref{GAL-mixture}). The integrand
of \linebreak
$\int_{\mathbb{R}^{+}}
\text{N}(y \mid \mu + \sigma \alpha s + \sigma A(p) z, \sigma^{2} B(p) z)
\, \text{Exp}(z \mid 1) \, \dd z$ can be recognized as the kernel of
a generalized inverse-Gaussian density. Therefore, integrating out $z$,
we obtain $f(y \mid p,\alpha,\mu,\sigma)=$
$\int_{\mathbb{R}^{+}}  p (1-p) \sigma^{-1}
\exp\lc - \sigma^{-1}  \lb p-I(y<\mu+\sigma\alpha s)\rb
[y - (\mu + \sigma \alpha s)] \rc \, \text{N}^{+}(s \mid 0,1) \,\dd s$.
This integral involves a normal density kernel, but care is needed
with the limits of integration which depend on the sign of $y-\mu$
and of $\alpha$. Combining the resulting expressions from all possible
cases, we obtain that for $\alpha\ne0$, the GAL density is given by
{\small
\begin{eqnarray}
    f(y \mid p,\alpha,\mu,\sigma)
            &=& 2\,\frac{p(1-p)}{\sigma}\,
                \left(\left[\Phi\left(\frac{y^\ast}{\alpha}
                -p_{\alpha_-}\alpha\right)-\Phi(- p_{\alpha_-}\alpha)\right]
                \exp\left\{-p_{\alpha_-} y^\ast
                +\frac{1}{2}(\,p_{\alpha_-}\alpha)^2\right\}
                I\left(\frac{y^\ast}{\alpha}>0\right)\right.\notag\\
            &~& \qquad\qquad\quad\left. +\,
                \Phi\left[ p_{\alpha_+}\alpha
                -\frac{y^\ast}{\alpha}
                 I\left(\frac{y^\ast}{\alpha}>0\right)
                \right]\exp\left\{-p_{\alpha_+} y^\ast
                +\frac{1}{2}\,(p_{\alpha_+}\alpha)^2\right\}\right)
                \label{eqn:gal0}
\end{eqnarray}}
where $y^\ast=(y-\mu)/\sigma$, $p_{\alpha_+}=p-I(\alpha>0)$,
$p_{\alpha_-}=p-I(\alpha<0)$, with $p\in(0,1)$.
The relatively complex form of the density in (\ref{eqn:gal0}) is not
an obstacle from a practical perspective, since its
hierarchical mixture representation facilitates study of model
properties and Markov chain Monte Carlo posterior simulation.

There is a direct link between the GAL
distribution and the $p_{0}$th quantile for any $p_{0} \in (0,1)$; note
that parameter $p$ no longer corresponds to the cumulative
probability at the quantile for $\alpha \ne 0$. When $\alpha>0$,
the distribution function of (\ref{eqn:gal0}) at $\mu$ is given by
$\int_{-\infty}^{\mu} f(y \mid p,\alpha,\mu,\sigma) \dd y=$
$2 p \Phi[(p-1)\alpha] \exp\left\{ (p-1)^{2}\alpha^{2}/2 \right\}$.
Hence, letting $\gamma=(1-p)\alpha$, the distribution function
becomes,
\begin{eqnarray*}
\int_{-\infty}^{\mu} f(y \mid p,\gamma,\mu,\sigma) \, \dd y =
p \, g(\gamma)
    &\quad& \mathrm{with}  \,\,\,\,\,\,\,\,\,
g(\gamma) = 2 \Phi(-|\gamma|) \exp(\gamma^{2}/2).
\end{eqnarray*}
We use $|\gamma|$ above, since this is the general form of $g(\gamma)$
that applies also in the $\alpha < 0$ case.

Note that, for $\gamma \in \mathbb{R}^{-}$,
$\dd g(\gamma)/ \dd \gamma =$
$2 h(\gamma) \exp(\gamma^{2}/2)$, where $h(\gamma)=$
$\phi(\gamma) + \gamma \Phi(\gamma)$.
The function $h(\gamma)$ is monotonically increasing in
$\mathbb{R}^{-}$, since $\dd h(\gamma) /\dd \gamma=$
$\Phi(\gamma)>0$. Moreover, $h(0)=$ $(2\pi)^{-1/2} > 0$, and
$\lim_{\gamma \rightarrow -\infty} h(\gamma) = 0$.
Therefore, $h(\gamma)>0$ for $\gamma \in \mathbb{R}^{-}$,
and thus $g(\gamma)$ is monotonically increasing in $\mathbb{R}^{-}$.
Since $g(\gamma)$ is an even function, it also obtains that it is
monotonically decreasing in $\mathbb{R}^{+}$.

Consider now setting
$\int_{-\infty}^{\mu} f(y \mid p,\gamma,\mu,\sigma) \, \dd y=$
$p g(\gamma) = p_{0}$. Then, the fact that $g(\gamma)$ is decreasing
in $\mathbb{R}^{+}$ combined with $g(\gamma) > p_0$, imply that
for each $\gamma>0$ in the domain that respects the condition of
$p\in(0,1)$ and $\alpha>0$, there is a unique solution of $p$ that
ensures $\int_{-\infty}^{\mu} f(y \mid p,\gamma,\mu,\sigma) \, \dd y=$
$p_{0}$, and subsequently a unique $\alpha$ based on $\gamma=(1-p)\alpha$.
For $\alpha<0$, setting
$\int^{\infty}_{\mu} f(y \mid p,\gamma,\mu,\sigma) \, \dd y=$ $1 - p_{0}$ and
letting $\gamma=$ $p\alpha$ leads to the same argument.

\begin{figure}[t!]
    \centering
    \includegraphics[height=3.5in,width=6.15in]{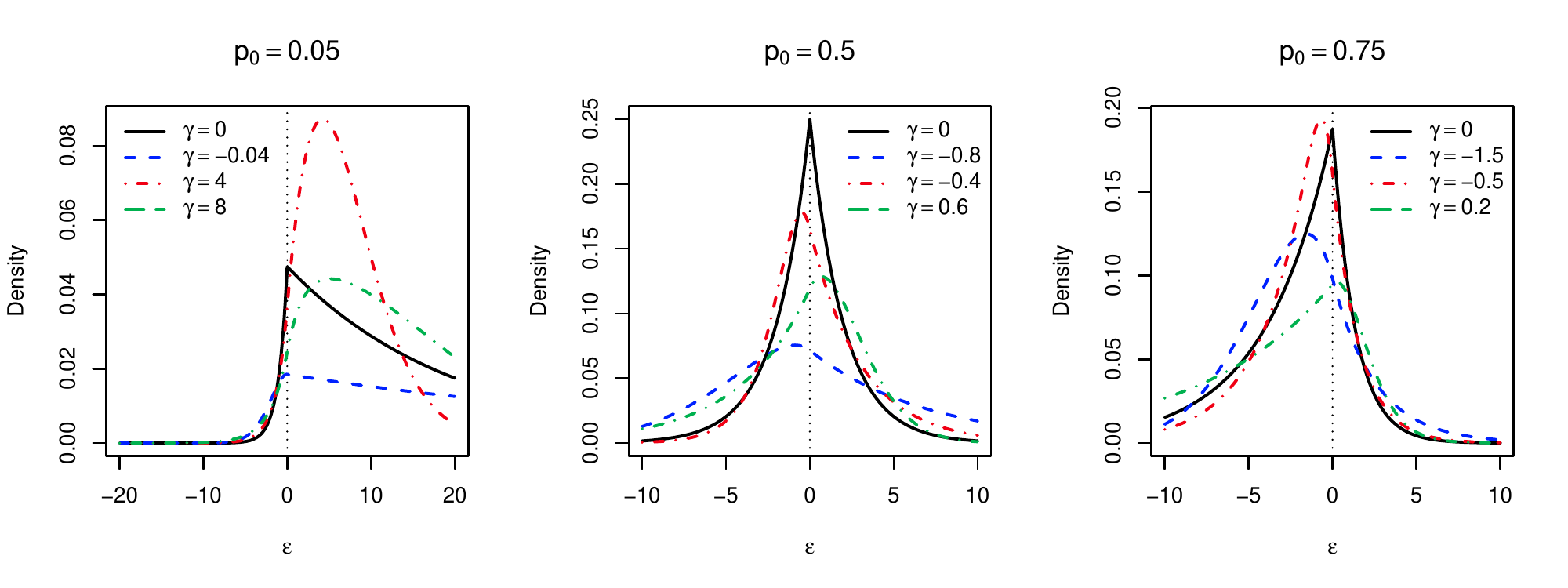}\vspace{-1em}
    \caption{Density function of quantile-fixed
      generalized asymmetric Laplace distribution with
$\mu=0$, $\sigma=1$ and different values of $\gamma$,
for $p_0=0.05$, $0.5$ and $0.75$. In all cases, the solid line
corresponds to the asymmetric Laplace density ($\gamma=0$).}
    \label{fig:density}
\end{figure}

The above connection between $(p_{0},\gamma)$ and $(p,\alpha)$ suggests that
by reparameterization with desired $p_{0}$ and
$\gamma=[I(\alpha>0)-p]|\alpha|$, we can derive a new family of
distributions with the percentile for fixed $p_{0}$ given by $\mu$,
and with an additional shape parameter $\gamma$. For $\gamma \ne 0$,
the density,  $f_{p_0}(y \mid \gamma,\mu,\sigma)$, of such
quantile-fixed GAL distribution is
{\small
\begin{eqnarray}
    &~& 2\,\frac{p(1-p)}{\sigma} \left(\left\{\Phi
        \left( -\frac{p_{\gamma_+} y^\ast}{|\gamma|}
        +\frac{p_{\gamma_-}}{p_{\gamma_+}}|\gamma|\right)
        -\Phi\left(\frac{p_{\gamma_-}}{p_{\gamma_+}}|\gamma|
        \right)\right\}\exp\left\{-p_{\gamma_-}
        y^\ast+\frac{\gamma^2}{2}\left(\frac{p_{\gamma_-}}
        {p_{\gamma_+}}\right)^2 \right\}
        I\left(\frac{y^\ast}{\gamma}>0\right)\right.\notag\\
    &~& \qquad \qquad \qquad \qquad \qquad \qquad \qquad\qquad \left. + \,
        \Phi\left[-|\gamma|
        +\frac{p_{\gamma_+} y^\ast}{|\gamma|}
         I\left(\frac{y^\ast}{\gamma}>0\right)\right]
        \exp\left\{-p_{\gamma_+} y^\ast+
        \frac{\gamma^2}{2}\right\}  \right)  
        \label{eqn:gal_p0}
\end{eqnarray}}
where $p \equiv p(\gamma,p_{0})=$
$I(\gamma < 0) + \{ [p_{0} - I(\gamma < 0)]/g(\gamma) \}$,
$p_{\gamma_+}=$ $p-I(\gamma>0)$, $p_{\gamma_-}=$ $p-I(\gamma<0)$,
and $y^\ast=$ $(y-\mu)/\sigma$. Parameter $\gamma$
has bounded support over interval $(L,U)$, where $L$ is the
negative root of $g(\gamma)=1-p_0$ and $U$ is the positive root of
$g(\gamma)=p_0$. For instance, $\gamma$ takes values in
$(-0.07,15.90)$, $(-1.09,1.09)$ and $(-2.90,0.39)$ when $p_{0}=0.05$,
$p_{0}=0.5$ and $p_{0}=0.75$, respectively.
When $\gamma=0$, the density reduces to the AL
density, which is also a limiting case of (\ref{eqn:gal_p0}).
The density function is continuous for all possible $\gamma$ values.

The quantile-fixed GAL distribution has
three parameters, $\mu$, $\sigma$ and $\gamma$. Note that $Y$
has density $f_{p_0}(\cdot \mid \gamma,\mu,\sigma)$ if and only if
$(Y-\mu)/\sigma$ has density $f_{p_0}(\cdot \mid \gamma,0,1)$.
Hence, similarly to the AL distribution, $\mu$ is a
location parameter and $\sigma$ is a scale parameter. The new
shape parameter $\gamma$ enables the extension relative to the
quantile-fixed AL distribution. As demonstrated
in Figure \ref{fig:density}, $\gamma$ controls skewness and tail behaviour,
allowing for both left and right skewness when the median is fixed, as
well as for both heavier and lighter tails than the asymmetric
Laplace, the difference being particularly emphatic for extreme percentiles.
Moreover, as $\gamma$ varies, the mode is no longer held fixed at
$\mu$; it is less than $\mu$ when $\gamma<0$ and greater than $\mu$
when $\gamma>0$. The above attributes render the proposed distribution
substantially more flexible than the AL distribution.

Finally, we note that parameter $\gamma$ satisfies likelihood
identifiability. Consider the location-scale standardized
density, $f_{p_0}(\cdot \mid \gamma,0,1)$,
which is effectively the model for the errors in quantile
regression. Then, assume $f_{p_0}(y \mid \gamma_{1},0,1)=$
$f_{p_0}(y \mid \gamma_{2},0,1)$, for all $y \in \mathbb{R}$. Given
that parameter $\gamma$ controls the mode of the density, this implies
that $\gamma_{1}$ and $\gamma_{2}$ must have the same sign.
Working with either of the two cases (that is, $\gamma_{1}>0$ and
$\gamma_{2}>0$ or $\gamma_{1}<0$ and $\gamma_{2}<0$)
in expression (\ref{eqn:gal_p0}),
we arrive at $g(\gamma_{1})=$ $g(\gamma_{2})$, which, based on the
monotonicity of function $g(\cdot)$, implies $\gamma_{1}=\gamma_{2}$.

\section{Bayesian quantile regression with GAL errors}

\subsection{Inference for linear quantile regression}
\label{BQR}

Consider continuous responses $y_{i}$ and the associated covariate
vectors $\bm{x}_{i}$, for $i=1,\ldots, n$. The linear quantile
regression model is set up as $y_{i}=$ $\bm{x}_i^T \bm{\beta}+\epsilon_i$, where the
$\epsilon_i$ arise independently from a quantile-fixed GAL distribution with
$\int_{-\infty}^{0} f_{p_0}(\epsilon \mid \gamma,0,\sigma) \dd \epsilon=$
$p_{0}$. Owing to the mixture representation of the new distribution,
the model for the data can be expressed hierarchically as follows
\begin{eqnarray}
\label{hierarchical-model}
    y_i \mid \bm{\beta},\gamma,\sigma,z_i,s_i  & \stackrel{ind.} \sim &
\text{N}(y_{i} \mid \bm{x}_i^T \bm{\beta} + \sigma C|\gamma| s_{i}
            +\sigma A z_{i}, \sigma^{2} B z_{i}), \,\,\, i=1,...,n
            \notag \\
z_{i}, s_{i} & \stackrel{ind.} \sim & \text{Exp}(z_{i} \mid 1) \, \text{N}^{+}(s_{i} \mid 0,1),
\,\,\, i=1,...,n
\end{eqnarray}
where $C=[I(\gamma>0)-p]^{-1}$, and $A$ and $B$ are the functions of $p$
given in (\ref{AL-mixture}). Since $p$ is a function of $\gamma$ and $p_{0}$,
$A$, $B$ and $C$ are all functions of parameter $\gamma$.
The Bayesian model is completed with priors for $\bm{\beta}$, $\sigma$
and $\gamma$. Here, we assume a normal prior $\text{N}(\bm{m}_{0},\Sigma_0)$
for $\bm{\beta}$ and an inverse-gamma prior $\text{IG}(a_{\sigma},b_{\sigma})$
for $\sigma$, with mean $b_{\sigma}/(a_{\sigma}-1)$ provided $a_{\sigma}>1$.
For any specified $p_{0}$,  $\gamma$ is defined over an interval $(L,U)$ with
fixed finite endpoints, and thus a natural prior for $\gamma$ is given
by a rescaled Beta distribution, with the uniform distribution
available as a default choice.

The augmented posterior distribution, which includes the $z_{i}$ and
the $s_{i}$, can be explored via a Markov chain Monte Carlo algorithm
based on Gibbs sampling updates for all parameters other than $\gamma$.
As in \cite{KozuKoba2011}, we set $v_i=$ $\sigma z_i$, $i=1,\ldots, n$.
Then, the posterior simulation method is based on the following updates.
\begin{itemize}
    \item[1.] Sample $\bm{\beta}$ from
      $\text{N}(\bm{m}^{\ast},\Sigma^{\ast})$, with covariance matrix
$\Sigma^\ast=$ $[\Sigma_0^{-1} + \sum_{i=1}^n \bm{x}_i \bm{x}_i^T/(B \sigma v_i)]^{-1}$
and mean vector $\bm{m}^{\ast}=$ $\Sigma^\ast \{ \Sigma_0^{-1}\bm{m}_{0} +  
\sum_{i=1}^n \bm{x}_i [y_i-(\sigma C|\gamma|s_i+Av_i)]/(B\sigma v_i) \}$.
    \item[2.] For each $i=1,...,n$, sample $v_i$ from a generalized inverse-Gaussian
      distribution, $\text{GIG}(0.5,a_i,b_i)$, where
      $a_i=[y_i-(\bm{x}_i^T \bm{\beta} +\sigma C|\gamma|s_i)]^2/(B\sigma)$ and
      $b_i=2/\sigma + A^{2}/(B\sigma)$, with density given by
$\text{GIG}(x \mid \nu,a,b) \propto$ $ x^{\nu-1} \exp\{-0.5(a/x+bx)\}$.
    \item[3.] For each $i=1,...,n$, sample $s_i$ from a normal 
$\text{N}(\mu_{s_i},\sigma_{s_i}^{2})$ distribution truncated on $\mathbb{R}^{+}$, 
where $\sigma_{s_i}^{2}=$ $[(C\gamma)^2\sigma/(Bv_i)+1]^{-1}$ and $\mu_{s_i}=$ 
$\sigma_{s_i}^{2} C |\gamma| [y_i-(\bm{x}_i^T \bm{\beta} + A v_i)]/(Bv_i)$.
    \item[4.] Sample $\sigma$ from a $\text{GIG}(\nu,c,d)$ distribution,
where $\nu=$ $-(a_{\sigma} + 1.5n)$,
$c=$ $2b_{\sigma} + 2\sum_{i=1}^n v_i+\sum_{i=1}^n[y_i-(\bm{x}_i^T\bm{\beta} +Av_i)]^2/(Bv_i)$,
and $d=\sum_{i=1}^n(C\gamma s_i)^2/(Bv_i)$.
    \item[5.] Update $\gamma$ with a Metropolis-Hasting step, using
a normal proposal distribution on the logit scale over $(L,U)$.
\end{itemize}

Based on the hierarchical model structure, the posterior predictive error
density can be expressed as $p(\epsilon \mid \text{data})=$
$\int \text{N}(\epsilon \mid \sigma C|\gamma|s+\sigma A z,\sigma^{2} B z)
        \, \text{Exp}(z \mid 1)\notag \, \text{N}^{+}(s \mid 0,1) \,
        \pi(\gamma,\sigma\,|\,\mathrm{data})\,
        \dd s\,\dd z \, \dd \gamma \, \dd \sigma$,
and thus estimated through Monte Carlo integration,
using the posterior samples of $(\gamma,\sigma)$.

\subsection{Quantile regression with regularization}
\label{lasso}

Since the GAL distribution is constructed through modifying the
mixture representation of the AL distribution, it retains some of the
interesting properties of the AL distribution. In particular, 
working with the hierarchical representation of the GAL distribution,
we are able to retrieve an extended version of the check loss function
which corresponds to asymmetric Laplace errors.

Consider the collapsed posterior distribution,
$\pi(\bm{\beta},\gamma,\sigma,s_{1},...,s_{n} \mid \text{data})$,
that arises from (\ref{hierarchical-model}) by marginalizing over the $z_{i}$.
Then, the corresponding posterior full conditional for $\bm{\beta}$ can be expressed
as
\[
\pi(\bm{\beta} \mid \gamma,\sigma,s_{1},...,s_{n},\text{data}) \propto \pi(\bm{\beta})
\exp \left\{ -\frac{1}{\sigma} \sum_{i=1}^{n}
\rho_{p}(y_{i} - \bm{x}^{T}_{i} \bm{\beta} - \sigma H(\gamma) s_{i}) \right\}
\]
where $\pi(\bm{\beta})$ is the prior density for $\bm{\beta}$, $H(\gamma)=$
$\gamma g(\gamma) / \{ g(\gamma) - |p_{0} - I(\gamma < 0)| \}$, and
$p=$ $I(\gamma < 0) + \{ [p_{0} - I(\gamma < 0)]/g(\gamma) \}$,
with $p_{0}$ the probability associated with the specified quantile
modeled through $\bm{x}^{T}_{i} \bm{\beta}$.
Hence, ignoring the prior contribution, finding the mode of the
posterior full conditional for $\bm{\beta}$ is equivalent to minimizing
with respect to $\bm{\beta}$ the adjusted loss function
$\sum_{i=1}^{n} \rho_{p}(y_{i} - \bm{x}^{T}_{i} \bm{\beta} - \sigma H(\gamma) s_{i})$;
note that in the special case with asymmetric Laplace errors, that is, for
$\gamma=0$, this reduces to the check loss function with $p=$ $p_{0}$.

Based on the above structure, the positive-valued latent variables $s_{i}$
can be viewed as response-specific weights that are adjusted by real-valued
coefficient $H(\gamma)$, which is fully specified through the shape parameter
$\gamma$. The result is the real-valued, response-specific terms
$\sigma H(\gamma) s_{i}$, which reflect on the estimation of $\bm{\beta}$
the effect of outlying observations relative to the AL
distribution. A promising direction to further explore this structure
is in the context of variable selection. For instance, \cite{LXL2010}
study connections between different versions of
regularized quantile regression and different priors for $\bm{\beta}$,
working with asymmetric Laplace errors. The main example
is lasso regularized quantile regression, which can be connected to the
Bayesian asymmetric Laplace error model through a hierarchical Laplace
prior for $\bm{\beta}$. We consider this prior below extending the AL
error distribution to the proposed GAL distribution.
The perspective we offer may be useful, since
it can be used to explore regularization adjusting the loss function,
through the response distribution, in addition to the penalty term,
through the prior for the regression coefficients.

Here, we denote by $\bm{\beta}$ the $d$-dimensional vector of regression
coefficients excluding the intercept $\beta_{0}$. Then, the Laplace
conditional prior structure for $\bm{\beta}$ is given by
\[
    \pi(\bm{\beta} \mid \sigma,\lambda)
    = \prod_{k=1}^d\frac{\lambda}{2 \sigma}\exp\lc- \frac{\lambda}{\sigma}|\beta_k|\rc
    = \prod_{k=1}^d\int_{\mathbb{R}^{+}} \frac{1}{\sqrt{2\pi\omega_k}}
        \exp\lc-\frac{\beta_k^2}{2\omega_k}\rc\frac{\eta^2}{2}
        \exp\lc-\frac{\eta^2}{2}\omega_k\rc\,\dd \omega_k.
\]
The second expression above utilizes the normal scale mixture
representation for the Laplace distribution, which has been exploited
for posterior simulation in the context of lasso mean regression
\citep{ParkCase2008}. Moreover, to facilitate Markov chain Monte Carlo
sampling, we reparameterize in terms of $\eta=$ $\lambda/\sigma$
and place a gamma prior on $\eta^{2}$. The lasso regularized version
of model (\ref{hierarchical-model}) is completed with a normal prior
for $\beta_{0}$, and with the priors for the other parameters as given
in Section \ref{BQR}. The posterior simulation algorithm is the same with
the one described in Section \ref{BQR} with the exception of the
updates for the $\beta_{k}$, $k=1,...,d$, and for $\eta^{2}$. 
Using the mixture representation of the Laplace prior, each $\beta_{k}$
can be sampled from a normal distribution, whereas $\eta^{2}$ has a gamma 
posterior full conditional distribution.

\subsection{Tobit quantile regression}
\label{tobit}

Tobit regression offers a modeling strategy for problems involving range
constraints on the response variable \citep{Amem1984}. The standard Tobit
regression model can be viewed in the context of censored regression where the 
responses are left censored at a threshold $c$; without loss of generality, we take $c=0$. 
The responses can be written as $y_i=$ $\max\{0,y_i^\ast\}$, where $y_i$ are
the observed values and $y_i^\ast$ are latent if $y_i^\ast\le0$. In the context 
of quantile regression, \cite{YuStan2007} and \cite{KozuKoba2011}
applied the AL-based model to the latent responses $y_i^\ast$. Here, we 
consider the Tobit quantile regression setting with GAL errors.

Consider a data set of $n+k$ observations on covariates and 
associated responses $\bm{y}=$
$(\bm{y}^{\mathrm{o}},\bm{0})$, where $\bm{y}^{\mathrm{o}}=$ 
$(y_{1}^{\mathrm{o}},...,y_{n}^{\mathrm{o}})$ consists of positive-valued 
observed responses with the remaining $k$ responses censored
from below at $0$. Assuming the GAL distribution for the latent
responses, the likelihood can be expressed as
%
%
$\prod_{i=1}^{n} f_{p_0}(y_i^{\mathrm{o}} \mid \gamma,\bm{x}_{i}^{T} \bm{\beta},\sigma)
\prod_{j=1}^{k} \int_{-\infty}^{0} f_{p_0}(w \mid \gamma,\bm{x}_{n+j}^{T} \bm{\beta},\sigma)
\, \dd w$. 
Using data augmentation \citep{Chib1992}, let $\bm{w}=$ 
$(w_1,...,w_k)$ be the unobserved (latent) responses corresponding to the 
$k$ data points that are left censored at $0$. Then, using again the 
hierarchical representation of the GAL distribution, the joint posterior 
distribution that includes $\bm{w}$ can be written as 
\[
\begin{array}{l}
p(\bm{\beta},\gamma,\sigma,\{ s_{i} \},\{ v_{i} \},\bm{w} \mid \text{data})
\propto
\pi(\bm{\beta},\gamma,\sigma)
\prod_{i=1}^n \text{N}(y_i^{\mathrm{o}} \mid \bm{x}_{i}^{T} \bm{\beta} +
        \sigma C|\gamma|s_i+Av_i,\sigma Bv_i) \\
\prod_{j=1}^k \text{N}^{-}(w_j \mid \bm{x}_{n+j}^T \bm{\beta}+
        \sigma C|\gamma|s_{n+j}+Av_{n+j},\sigma Bv_{n+j}) 
\prod_{i=1}^{n+k}\,
        \text{Exp}(v_i \mid \sigma^{-1}) \, \text{N}^+(s_i \mid 0,1)
\end{array}
\]
where $\pi(\bm{\beta},\gamma,\sigma)$ denotes the prior for the model
parameters, and $v_i=$ $\sigma z_i$. Here, $\text{N}^-$ denotes a
truncated normal on $\mathbb R^-$, and $\text{Exp}(v\,|\,\sigma^{-1})$ an
exponential distribution with mean $\sigma$.

Regarding posterior inference, the posterior full conditional 
for each auxiliary variable $w_j$ is given by a truncated normal distribution. 
And, given the augmented data $(\bm{y}^{\mathrm{o}},\bm{w})$, 
the model parameters and the latent variables $\{ (v_{i},s_{i}): i=1,...,n+k \}$ 
can be sampled as before.
%
%
%

Although results are not reported here, we have tested the posterior
simulation algorithm on simulated data sets based on GAL errors, with $n=400$ observations
and a censoring rate that ranged from $20\%$ to $40\%$. Under this
scenario, the posterior distributions successfully captured the true
values of all parameters in their 95\% credible intervals.
%
%

\section{Simulation study}

Here, we present results from a simulation study designed to compare 
the lasso regularized quantile regression models with AL and GAL errors.
We follow a standard simulation setting from 
the literature regarding the linear regression component 
\citep{Tibs1996,ZouYuan2008,LXL2010}, varying the extent of sparsity in the true
$\bm{\beta}$ vector. For the underlying data-generating error distributions,
we consider four scenarios with different types of skewness and tail behavior.
For model comparison, we evaluate the accuracy in variable selection,
inference for the regression function, and posterior predictive
performance, using relevant assessment criteria. 
Overall, the GAL-based quantile regression model performs better in
variable selection and prediction accuracy and it is more robust to
non-standard error distributions, particularly for extreme quantiles. 
The two models yield comparable results in the case of median
regression.

\subsection{Simulation settings}

We consider synthetic data generated from linear quantile regression settings,
with $p_0=0.05$, $0.25$ and $0.5$ to study model performance
for both extreme and more central percentiles. The rows of the design
matrix were generated independently from an 8-dimensional
normal distribution with zero mean vector and covariance matrix with
elements $0.5^{|i-j|}$, for $1\le i,j\le 8$. We present detailed
results from a relatively sparse case for the vector of regression
coefficients, $\bm{\beta}=$ $(3,1.5,0,0, 2,0,0,0)$. In Section \ref{simulation-results},
we briefly discuss results form two other scenarios for $\bm{\beta}$
corresponding to a dense and a very sparse case.

Data were simulated under four different error distributions: 
\begin{itemize}
    \item $\mbox{N}(\mu,9)$, with $\mu$ chosen such that the $p_0$th quantile is 0.
    \item $\mbox{Laplace}(\mu,3)$, with $\mu$ chosen such that the $p_0$th quantile is 0.
    \item $0.1\mbox{N}(\mu,1)
        +0.9\mbox{N}(\mu+1,5)$, with $\mu$ chosen such that the $p_0$th quantile is 0.
    \item $\mbox{Log-transformed generalized Pareto}(\sigma,\xi)$,
      with $\xi=3$ and $\sigma$ chosen such
that the $p_0$th quantile is 0. To generate the errors, we first
sample from a generalized Pareto distribution, then take the
logarithm. Based on the parameterization in \cite{EmbrKlup1997}, 
the density function of the errors is given by
$f(\epsilon\,|\,\sigma,\xi) =$
$\sigma^{-1} \{ 1 + \xi  \sigma^{-1} \exp(\epsilon) \}^{-(1+\xi^{-1})}
\exp(\epsilon)$, for $\epsilon\in\mathbb{R}$.
%
%
\end{itemize}
The normal and Laplace error distributions are symmetric about zero
under median regression. The parameters of the two-component normal
mixture are selected such that the resulting error distribution is skewed.
Finally, the log-transformed generalized Pareto distribution is included
to study model performance under an error density which is both skewed and
does not have exponential tails.

For each setting of the simulation study, we generated 100 data sets,
each with $n=100$ observations for training the models and another $N=100$ 
for testing predictions.

\subsection{Criteria for comparison}
\label{sec:lasso_comparison}

We consider a number of criteria to assess different aspects of model 
performance. Since Bayesian lasso regression only shrinks the
covariate effects, we consider a threshold on the effect size for the
purpose of variable selection. Following 
\cite{HotiSill2006}, we calculate the standardized effects as
$\beta_j^\ast=$ $(s_{x_j}/s_y) \beta_{j}$, $j=1,\ldots,d$,
where $s_{x_j}$ is the standard deviation of predictor $x_j$ and
$s_y$ is the standard deviation of the response. For each
posterior sample, if the standardized effect is greater than 0.1 in
absolute value, we consider the predictor as included. We count the
number of correct inclusion and exclusions (CIE) in the posterior 
sample and divide it by $d$ to normalize it to a number between 0 and 1. 
By averaging over all the posterior samples, we obtain the mean 
standardized CIE for each simulated data set.

To assess predictive performance for the regression function, we
calculate the mean check loss on the $N$ test data points, 
defined as: $\mathrm{MCL}=$ $N^{-1} \sum_{i=1}^{N} 
\rho_{p_0}(\bm{x}_{i}^{T} \bm{\beta}^{*} - \bm{x}_{i}^{T} \bm{\beta})$, where
$\bm{\beta}^{*}$ is the posterior mean estimate from the training data. 
The mean check loss resembles the standard mean squared error 
criterion, which is commonly used for evaluating prediction with cross-validation.

Finally to assess model fitting taking into account predictive 
uncertainty, we apply the posterior predictive loss
criterion from \cite{GelfGhos1998}. This criterion favors the model
$\mathcal{M}$ that minimizes $D_m(\mathcal{M})=$ $P(\mathcal{M}) + \{
m/(m+1) \} G(\mathcal{M})$, where $G(\mathcal{M})=$ $\sum_{i=1}^{n} \{
y_i -\text{E}^{\mathcal{M}}(y^{*}_{i} \mid \text{data}) \}^2$ is a
goodness-of-fit term, and $P(\mathcal{M})=$ $\sum_{i=1}^{n}
\text{var}^{\mathcal{M}}(y^{*}_{i} \mid \text{data})$ is a penalty
term for model complexity. Here, $m\ge0$, and
$\text{E}^{\mathcal{M}}(y^{*}_{i} \mid \text{data})$ and
$\text{var}^{\mathcal{M}}(y^{*}_{i} \mid \text{data})$ are the mean
and variance under model $\mathcal{M}$ of the posterior predictive
distribution for replicated response $y^{*}_{i}$ with corresponding
covariate $\bm{x}_{i}$. We also consider the generalized version of the
criterion based on the check loss function, under which $D(\mathcal{M})=$ 
$\sum_{i=1}^{n} \text{E}^{\mathcal{M}}(\rho_{p_0}(y_{i} - y^{*}_{i}) \mid \text{data})$. For this
generalized criterion, the goodness-of-fit term can be defined by
$G(\mathcal{M})=$ $\sum_{i=1}^{n} \rho_{p_0}( y_{i}
-\text{E}^{\mathcal{M}}(y^{*}_{i} \mid \text{data}) )$ and the
penalty term by $P(\mathcal{M})=$ $D(\mathcal{M})-G(\mathcal{M})$,
since the check loss function $L(y,a) \equiv \rho_{p_0}(y-a)=$ $(y-a)p_{0}
- (y-a)I(y < a)$ is convex in $y$, and thus $P(\mathcal{M}) \geq 0$; 
see \cite{GelfGhos1998} for details on defining the model comparison 
criterion under loss functions different from quadratic loss.

\subsection{Results}
\label{simulation-results}

We used the same hierarchical Laplace prior for $\bm{\beta}$ 
under the AL and GAL models, with a gamma prior for $\eta^{2}$
with prior mean $1$ and variance $10$. Such prior specification is relatively 
non-informative in the sense that it does not favor shrinkage for the regression
coefficients, resulting in marginal prior densities for each $\beta_{k}$
that place substantial probability mass away from $0$. The shape parameter 
$\gamma$ of the GAL error distribution was assigned a uniform prior. 
Results under both models and for each simulated data set are based on 
5,000 posterior samples, obtained after discarding the first 50,000 iterations 
of the Markov chain Monte Carlo sampler and then retaining one every 20 iterations.

Within each simulation scenario, we summarize results from the 100
data sets using the median and standard deviation (SD) of the values
for the performance assessment criteria discussed in Section 
\ref{sec:lasso_comparison}.
Results are reported in Table \ref{tab:simulation1} through Table \ref{tab:simulation4},
where we use boldface to indicate the model supported by the particular
criterion under each setting.

Overall, the lasso regularized Bayesian quantile regression model
performs better under the GAL error distribution.
The GAL-based model includes/excludes correct regression coefficient values 
more often than the AL model for almost all combinations of $p_0$ and 
error distributions (Table \ref{tab:simulation1}). It also results in a
lower median mean check loss for the test data in most cases, demonstrating
better performance in the prediction of the regression function (Table \ref{tab:simulation2}).
Note that, for both types of assessment in Tables \ref{tab:simulation1}
and \ref{tab:simulation2}, the GAL-based model produces better results
across all error distributions for $p_{0}=0.05$, and, with the exception
of one case, when $p_{0}=0.25$. Results are generally more balanced in the 
median regression setting, although the GAL model fares better in all
cases for which the underlying error distribution is skewed.

\begin{table}[t!]
    \centering
    \begin{tabular}{c l c c c c}
    \hline
    & & \multicolumn{4}{c}{Error distribution}\\\cline{3-6}
    &&&&& log-transformed\\
    ~~$p_0$ & Model & ~~~~~~~~Normal~~~~~~~ & ~~~~~~~Laplace~~~~~~~~ & Normal mixture
            & generalized Pareto\\
    \hline
     ~~0.05&GAL&\textbf{0.848} (0.063)&\textbf{0.633} (0.083)&\textbf{0.911} (0.042)&\textbf{0.893} (0.052) \\
    &AL&0.746 (0.099)&0.534 (0.087)&0.817 (0.075)&0.840 (0.081) \\[4.5pt]
    ~~0.25&GAL&\textbf{0.851} (0.049)&\textbf{0.728} (0.060)&\textbf{0.918} (0.048)&0.896 (0.050) \\
    &AL&0.843 (0.069)&0.700 (0.068)&0.913 (0.060)&\textbf{0.900} (0.051) \\[4.5pt]
    ~~0.50&GAL&0.848 (0.052)&\textbf{0.738} (0.065)&\textbf{0.909} (0.049)&\textbf{0.897} (0.055) \\
    &AL&\textbf{0.850} (0.056)&0.737 (0.065)&0.905 (0.050)&0.870 (0.061) \\
    \hline
    \hline
    \end{tabular}
    \caption{Simulation study. Standardized number of correctly included/excluded predictors: median (SD).}
    \label{tab:simulation1}
\end{table}

\begin{table}[t!]
    \centering
    \begin{tabular}{c l c c c c}
    \hline
    & & \multicolumn{4}{c}{Error distribution}\\\cline{3-6}
    &&&&& log-transformed\\
    ~~$p_0$ & Model & ~~~~~~~~Normal~~~~~~~ & ~~~~~~~Laplace~~~~~~~~ & Normal mixture
            & generalized Pareto\\
    \hline
    ~~0.05&GAL&\textbf{0.340} (0.083)&\textbf{1.073} (0.391)&\textbf{0.224} (0.060)&\textbf{0.268} (0.081) \\
    &AL&0.523 (0.130)&1.709 (0.485)&0.375 (0.101)&0.388 (0.114) \\[4.5pt]
    ~~0.25&GAL&\textbf{0.325} (0.086)&\textbf{0.676} (0.199)&\textbf{0.225} (0.071)&\textbf{0.265} (0.080) \\
    &AL&0.360 (0.096)&0.778 (0.215)&0.257 (0.076)&0.274 (0.082) \\[4.5pt]
    ~~0.50&GAL&0.323 (0.092)&0.642 (0.208)&\textbf{0.235} (0.064)&\textbf{0.262} (0.081) \\
    &AL&\textbf{0.322} (0.095)&\textbf{0.624} (0.207)&0.237 (0.063)&0.294 (0.089) \\
    \hline
    \hline
    \end{tabular}
    \caption{Simulation study. Mean check loss based on the test data: median (SD).}
    \label{tab:simulation2}
\end{table}

For each simulation setting, 
Table \ref{tab:simulation3} includes the values for the posterior predictive loss
criterion with quadratic loss (under $m \rightarrow \infty$, such that
$D_{\infty}=$ $P + G$), and Table \ref{tab:simulation4} shows the
generalized criterion under check loss.
%
%
%
Both versions of the posterior predictive loss criterion support the
GAL model when $p_{0}=0.05$, with differences in values between the two
models that are substantially larger than for the other two values of $p_{0}$.
This reinforces the earlier findings on the potential benefits of the
GAL error distribution for extreme percentiles. With the exception of
one case under the check loss version of the criterion, the GAL-based
model is also favored when $p_{0}=0.25$, whereas results are more
mixed in the median regression case.

Finally, although detailed results are not reported here, the simulation study 
included two more settings for $\bm{\beta}$, a dense case with all $8$
regression coefficients equal to $0.85$, and a very sparse case with 
$\bm{\beta}=$ $(5,0,0,0,0,0,0,0)$. 
The conclusions were overall similar, in particular, the GAL model 
outperformed the AL model for essentially
all combinations of underlying error distribution and value of $p_{0}=0.05$
or $p_{0}=0.25$. Again, in the median regression case, the distinction
between the two models was less clear for the normal, Laplace and
normal mixture data-generating distributions, although the GAL model 
performed better under all criteria for the setting corresponding to 
the log-transformed generalized Pareto distribution.

\begin{table}[t!]
    \centering
    \begin{tabular}{c l l c c c c}
    \hline
    & & &\multicolumn{4}{c}{Error distribution}\\\cline{4-7}
    &&&&&& log-transformed\\
    $p_0$ & Model & Score & ~~~~Normal~~~ & ~~~Laplace~~~~ & Normal mixture  & generalized Pareto\\
    \hline
    ~~0.05&GAL& $P$&\textbf{1231} (193)&\textbf{9799} (2483)&\textbf{653} (112)&\textbf{1273} (270) \\
    && $G$  &\textbf{832} (126)&\textbf{7046} (1546)&\textbf{429} (71)&\textbf{1053} (267) \\
    &&$D_\infty$&\textbf{2092} (312)&\textbf{16860} (3839)~~&\textbf{1085} (181)&\textbf{2319} (531) \\[2pt]
    &AL&$P$&3359 (799)&30308 (10763)&1839 (405)&2782 (664) \\
    &&$G$&952 (165)&8659 (2304)&534 (93)&1168 (279) \\
    &&$D_\infty$&4357 (933)&38766 (12676)&2398 (487)&4020 (873) \\[4.5pt]
    ~~0.25&GAL&$P$&\textbf{1085} (206)&\textbf{6977} (1607)&\textbf{608} (95)&\textbf{1445} (273) \\
    &&$G$&\textbf{~830} (146)&\textbf{6897} (1606)&\textbf{444} (66)&\textbf{1105} (264) \\
    &&$D_\infty$&\textbf{1882} (343)&\textbf{13884} (3115)&\textbf{1055} (154)&\textbf{2552} (511) \\[2pt]
    &AL&$P$&1630 (303)&11503 (2727)&884 (148)&1516 (260) \\
    &&$G$&865 (154)&7395 (1742)&464 (71)&1113 (263) \\
    &&$D_\infty$&2499 (448)&18916 (4349)&1352 (215)&2600 (487) \\[4.5pt]
    ~~0.50&GAL&$P$&1283 (205)&7600 (1676)&694 (97)&\textbf{1189} (217) \\
    &&$G$&\textbf{813} (132)&6459 (1509)&\textbf{424} (60)&\textbf{1089} (245) \\
    &&$D_\infty$&2111 (328)&14076 (3101)&1121 (152)&\textbf{2283} (415) \\[2pt]
    &AL&$P$&\textbf{1177} (191)&\textbf{7256} (1572)&\textbf{634} (87)&1318 (247) \\
    &&$G$&818 (134)&\textbf{6431} (1509)&426 (60)&1107 (255) \\
    &&$D_\infty$&\textbf{2008} (318)&\textbf{13667} (3019)&\textbf{1058} (143)&2415 (483) \\
    \hline
    \hline
    \end{tabular}
    \caption{Simulation study. Penalty term ($P$), goodness-of-fit term ($G$) and 
posterior predictive loss criterion ($D_\infty$) under quadratic loss: median (SD).}
    \label{tab:simulation3}
\end{table}

\begin{table}[t!]
    \centering
    \begin{tabular}{c l c c c c}
    \hline
    & & \multicolumn{4}{c}{Error distribution}\\\cline{3-6}
    &&&&& log-transformed\\
    ~~$p_0$ & Model & ~~~~~~~~Normal~~~~~~~ & ~~~~~~~Laplace~~~~~~~~ & Normal mixture
            & generalized Pareto\\
    \hline
    ~~0.05&GAL&\textbf{174.2} (13.6)&\textbf{507.0} (67.8)&\textbf{122.8} (11.3)&\textbf{178.5} (17.4) \\
    &AL&209.3 (21.7)&605.4 (70.8)&148.6 (17.3)&200.2 (20.5) \\[4.5pt]
    ~~0.25&GAL&\textbf{169.5} (15.9)&\textbf{443.9} (47.1)&\textbf{126.2} (9.5)&188.0 (17.5) \\
    &AL&178.0 (15.7)&451.4 (45.8)&129.0 (9.8)&\textbf{185.5} (17.3) \\[4.5pt]
    ~~0.50&GAL&175.7 (13.4)&444.7 (48.0)&127.4 (8.9)&\textbf{178.6} (16.1) \\
    &AL&\textbf{172.6} (13.4)&\textbf{438.5} (47.5)&\textbf{125.2} (8.7)&183.6 (18.1) \\
    \hline
    \hline
    \end{tabular}
    \caption{Simulation study. Posterior predictive loss criterion under check loss: median (SD).}
    \label{tab:simulation4}
\end{table}

\section{Data examples}

In this section, we consider three data examples to illustrate the
Bayesian quantile regression models developed 
in Sections \ref{BQR}, \ref{lasso}, and \ref{tobit}.
The main emphasis is on comparison of inference results between 
models based on the GAL distribution and those assuming 
an AL distribution for the errors.

We have implemented both models with priors for their parameters 
that result in essentially the same prior predictive error densities. 
The two models were applied with the same prior distributions for 
$\bm{\beta}$ and $\sigma$. More specifically, for 
the data sets of Sections \ref{igg} and \ref{labor-supply}, we used a 
$\text{N}(\bm{0},100I)$ prior for the vector of regression coefficients, and an 
$\text{IG}(2,2)$ prior for the scale parameter $\sigma$. For the data example
of Section \ref{boston}, we used a $\text{N}(0,100)$ prior for the intercept, 
and  the same conditional Laplace prior for the remaining regression coefficients 
with the simulation study (see Section \ref{simulation-results}). Finally,
a uniform prior was placed on the shape parameter $\gamma$ of the GAL 
error distribution. For all data examples, the posterior densities for
model parameters were fairly concentrated relative to the corresponding prior densities.
%
%

\subsection{Immunoglobulin-G data}
\label{igg}

We illustrate the proposed model, referred to as
model $\text{M}_{1}$, with a data set commonly used in additive quantile
regression; see, for instance, \cite{YuMoye2001}. The analysis focuses on
comparison with the simpler model based on asymmetric
Laplace errors, referred to as model $\text{M}_{0}$. The data set
contains the immunoglobulin-G concentration in grams per litre for
$n=298$ children aged between 6 months and 6 years. As in
earlier applications of quantile regression for these data, we use
a quadratic regression function
$\beta_{0} + \beta_{1} x+\beta_{2} x^{2}$ to model five quantiles, 
corresponding to $p_{0}=$ $0.05,\,0.25,\,0.5,\,0.75,\,0.95$, 
of immunoglobulin-G concentration against covariate age ($x$).

\begin{figure}[t!]
\centering
\begin{tabular}{ccc}
\includegraphics[height=2.5in,width=1.95in]{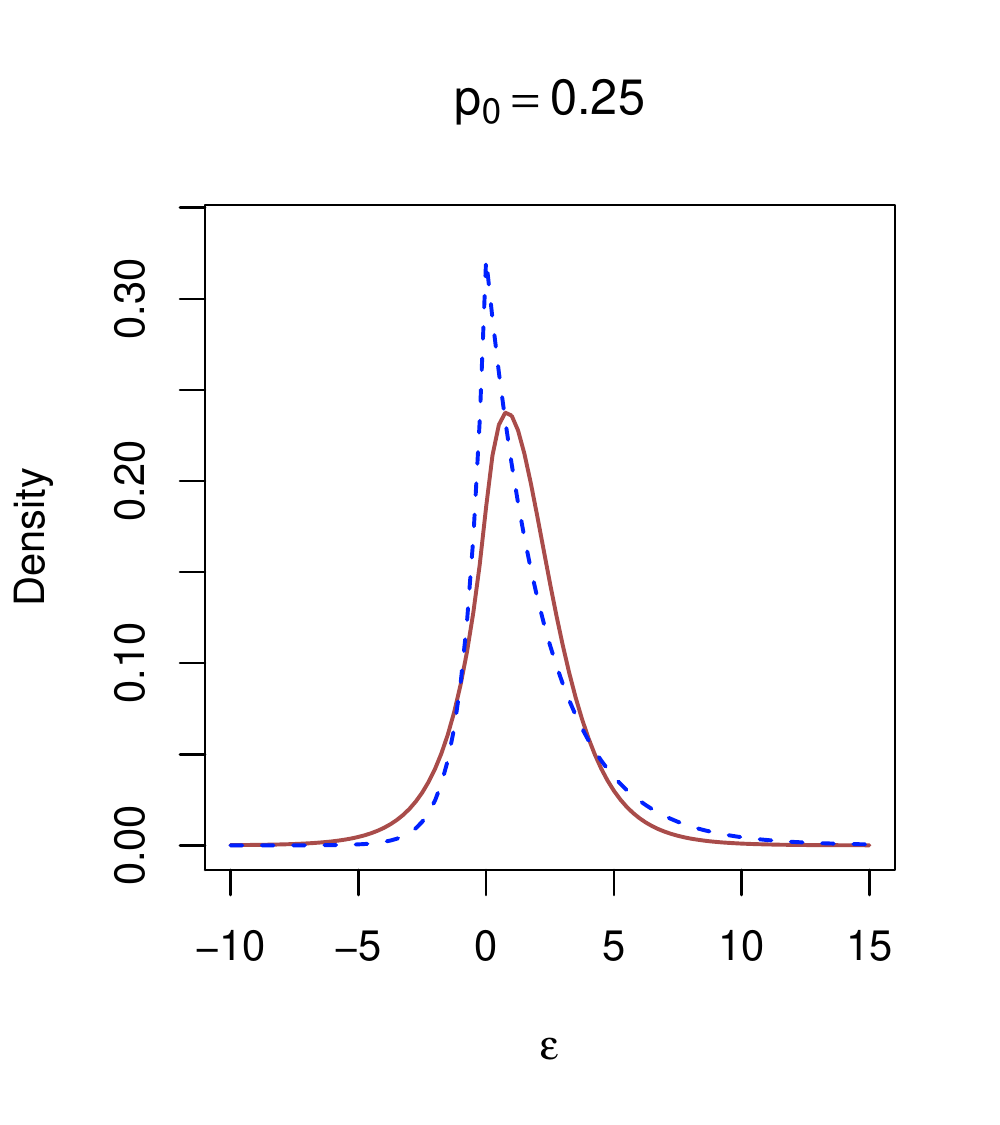} &
\includegraphics[height=2.5in,width=1.95in]{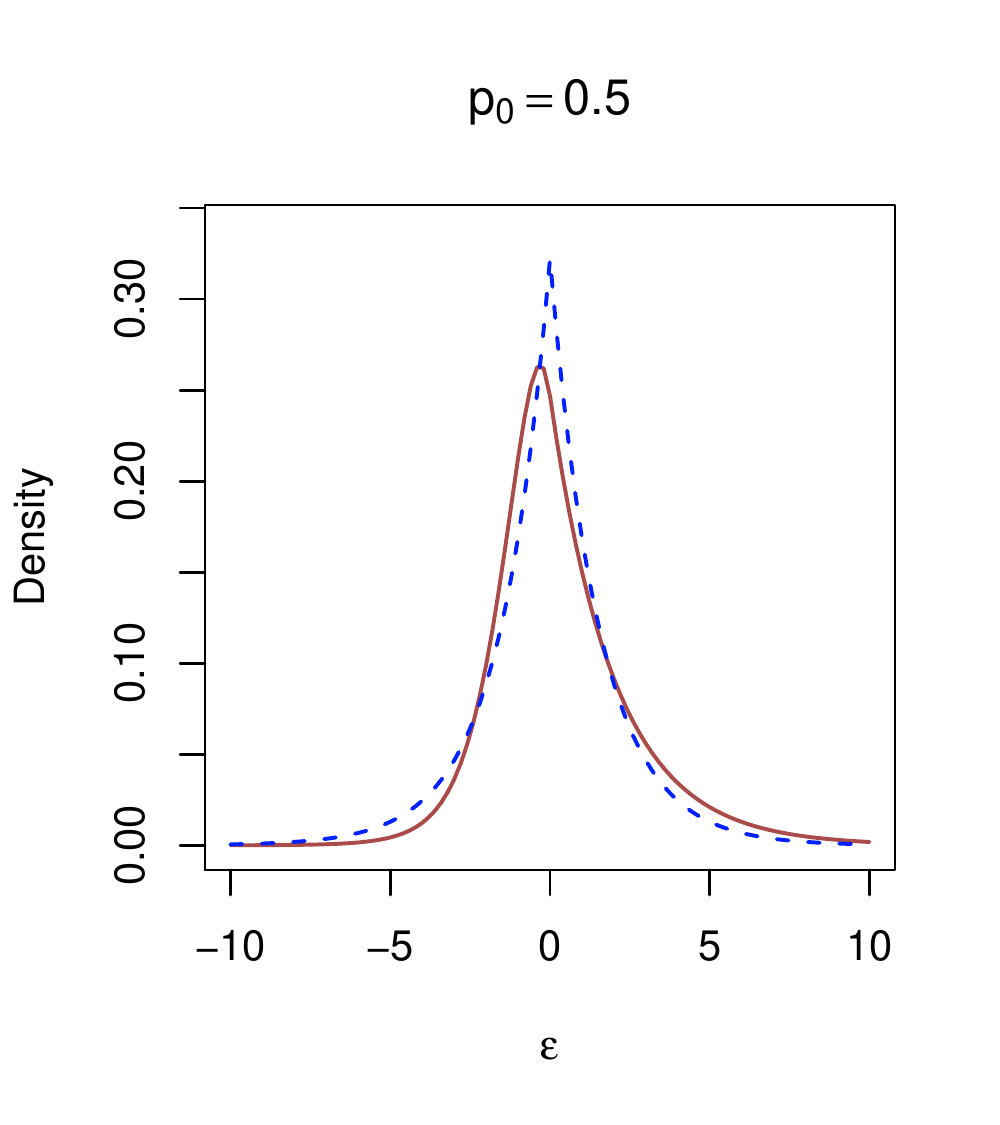} &
\includegraphics[height=2.5in,width=1.95in]{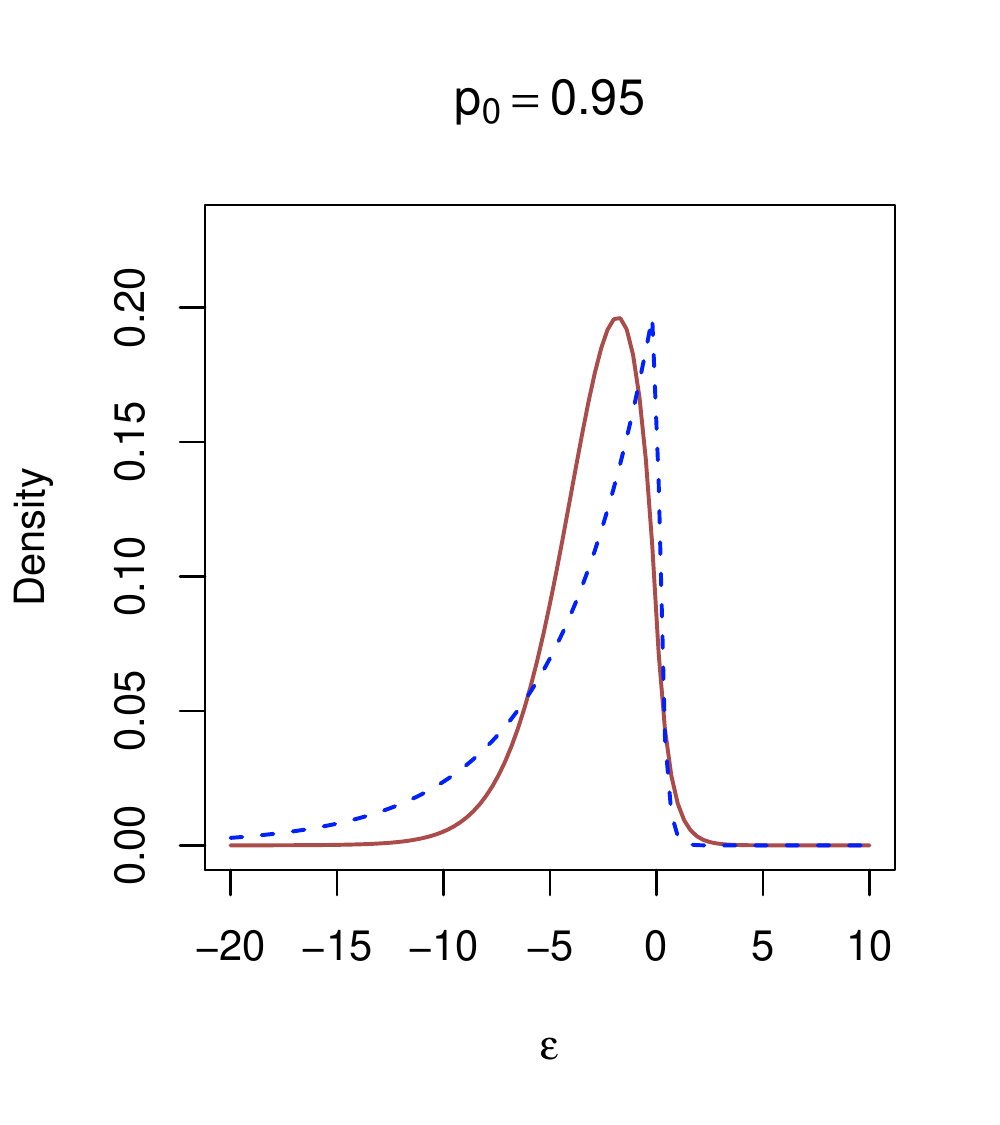} \\
\includegraphics[height=1.9in,width=1.95in]{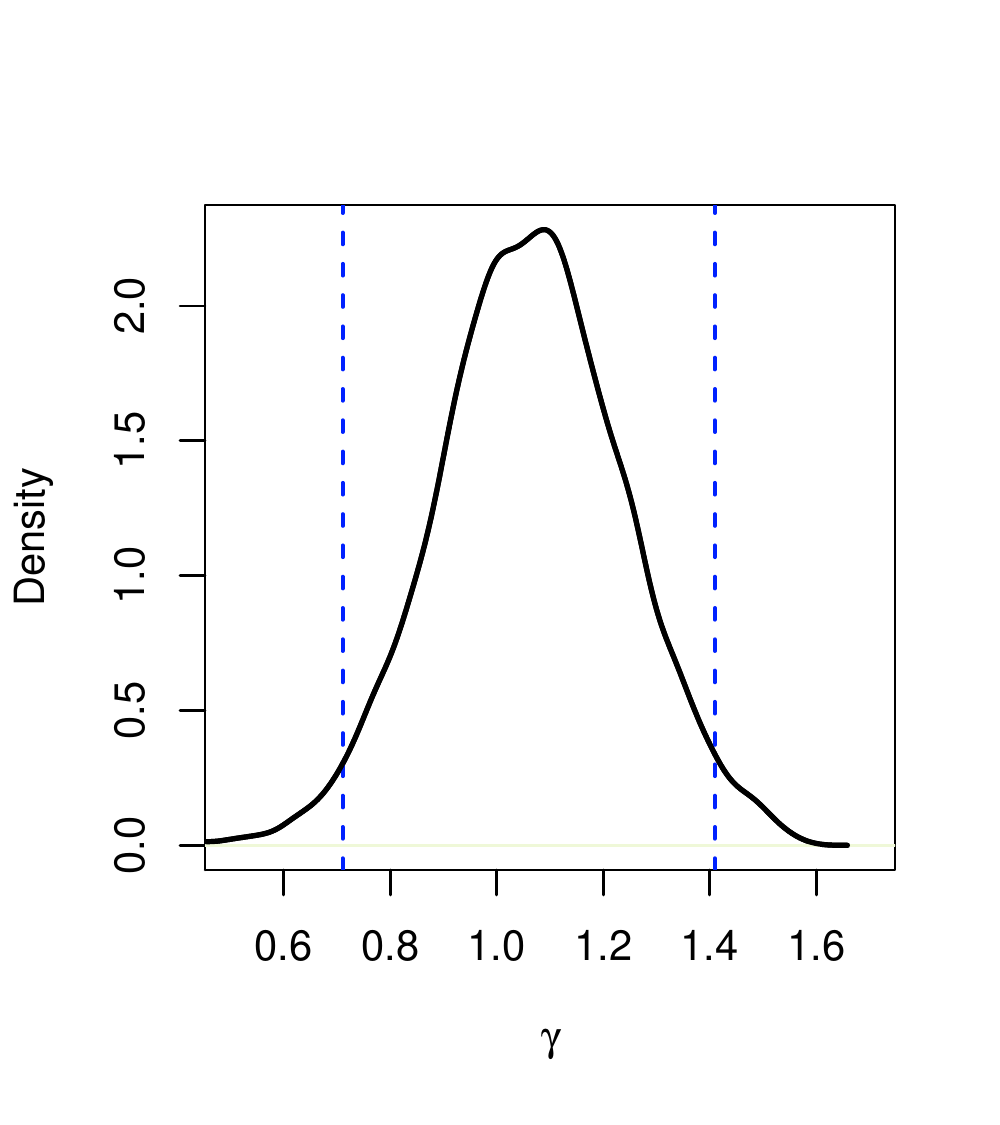} &
\includegraphics[height=1.9in,width=1.95in]{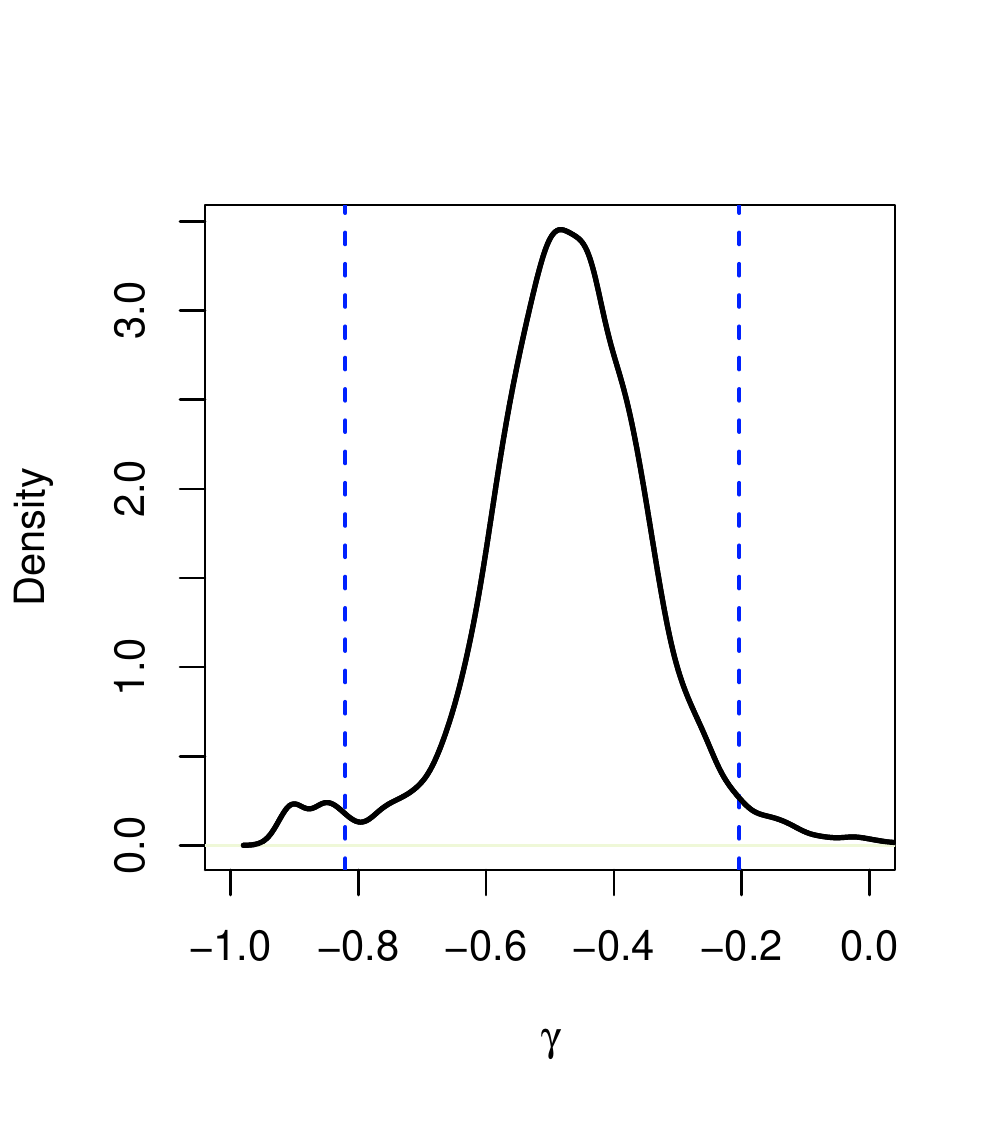} &
\includegraphics[height=1.9in,width=1.95in]{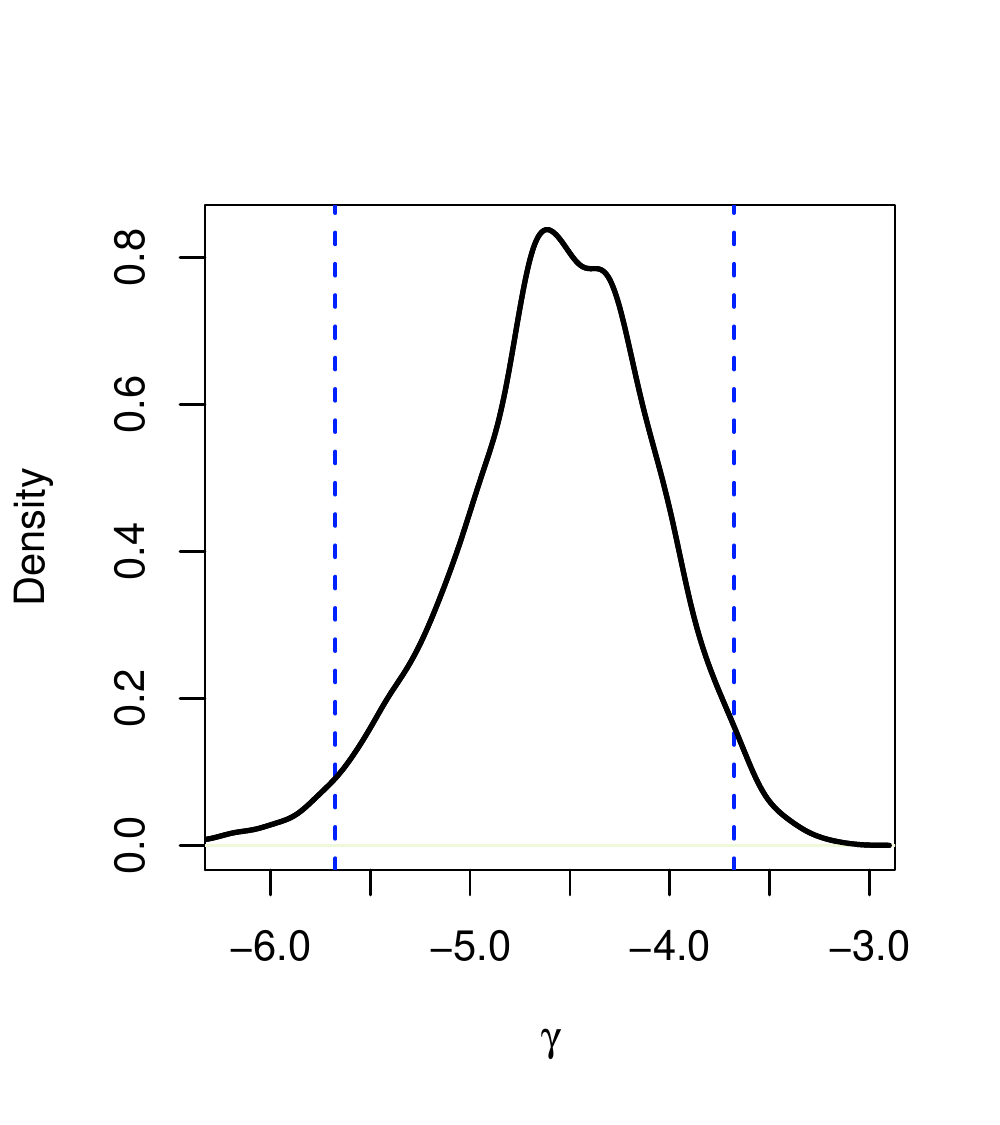}
\end{tabular}
    \caption{Immunoglobulin-G data. Inference results for
$p_0=0.25$, $0.5$ and $0.95$. Top row: posterior predictive error
densities under the asymmetric Laplace model (dashed lines)
and the generalized asymmetric Laplace model (solid lines). Bottom
row: posterior densities for parameter $\gamma$, with the vertical
lines corresponding to the endpoints of the 95\% credible interval.}
\label{fig:pred}
\end{figure}

\begin{figure}[t!]
\centering
\begin{tabular}{cc}
\includegraphics[width=.5\linewidth]{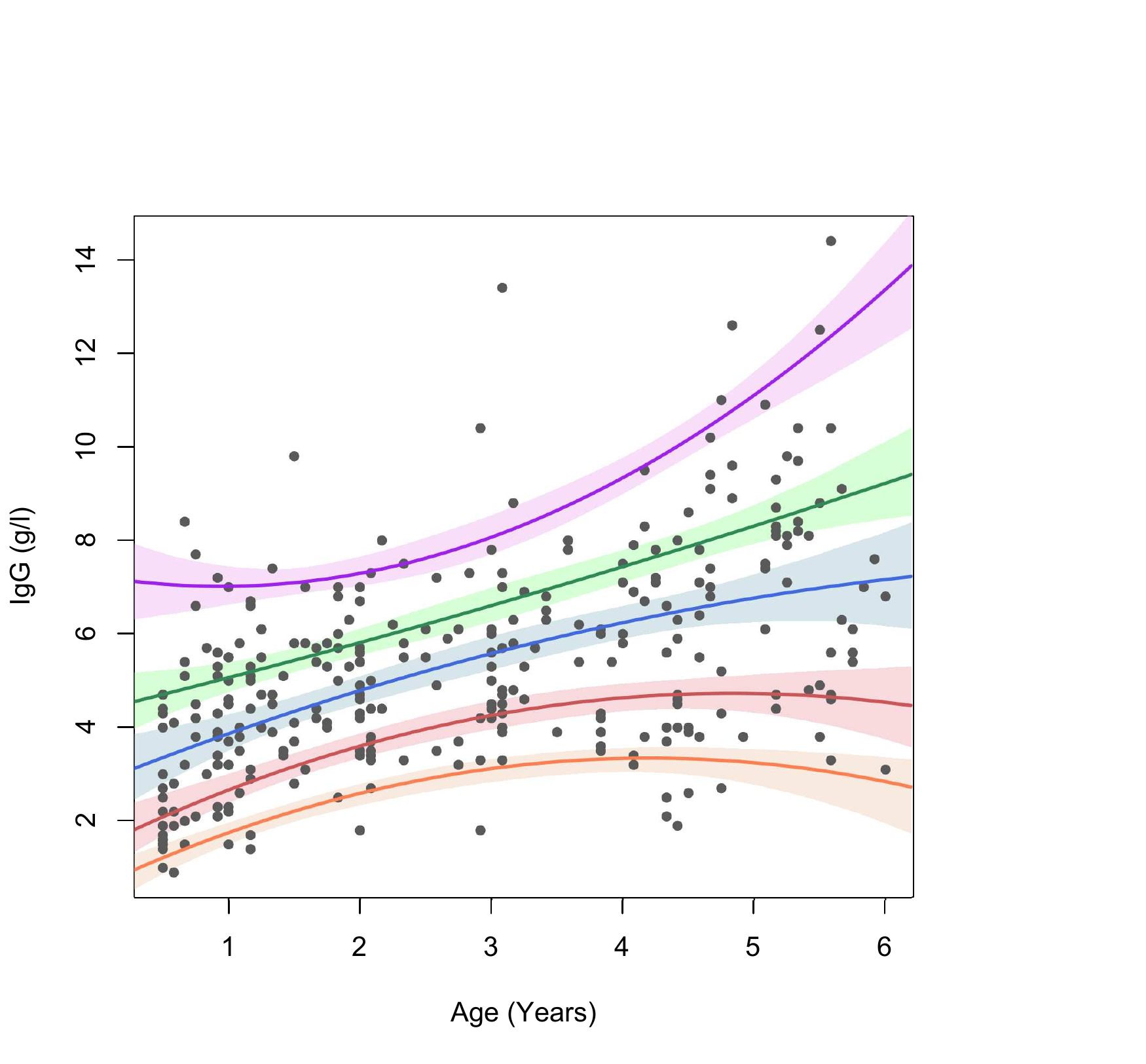} &
\includegraphics[width=.5\linewidth]{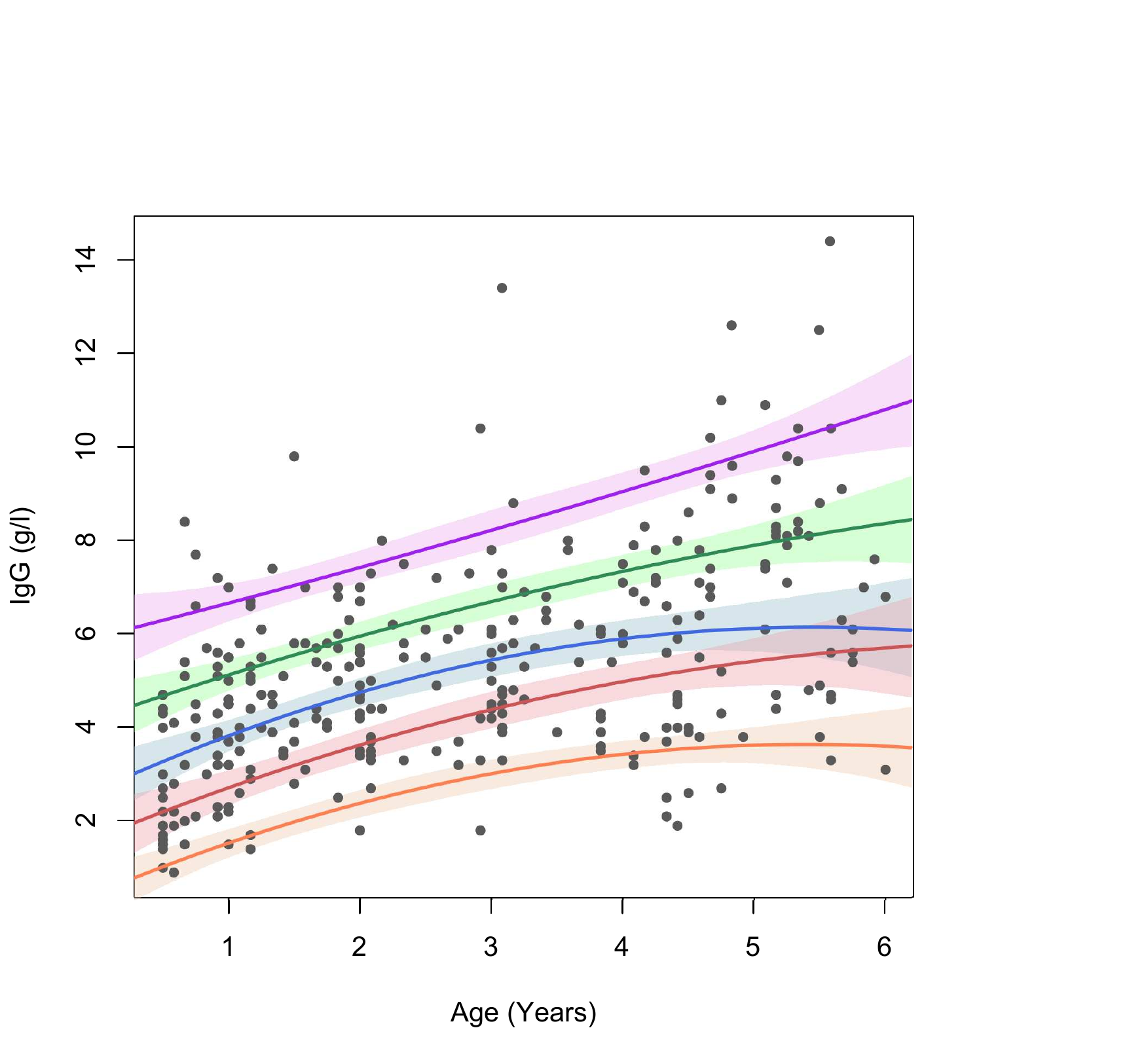}
\end{tabular}
    \caption{Immunoglobulin-G data. Posterior mean estimates and 95\%
credible bands for the quantile regression function $\beta_0+\beta_1x+\beta_2x^2$ 
against age ($x$), for $p_0=$ 0.05, 0.25, 0.50, 0.75 and 0.95. Left:
AL model. Right: GAL model.}
\label{fig:quantreg}
\end{figure}

The two models result in different posterior predictive error densities, 
especially for extreme percentiles; see Figure \ref{fig:pred}.
At $p_{0}=$ $0.95$, under the AL model,
both the shape and the skewness of the error distribution are predetermined
by $p_{0}$ and the mode is forced to be 0, resulting in a rigid heavy left tail. 
The effect of this overly dispersed tail can be observed in the
inference for the quantile regression function (Figure \ref{fig:quantreg}). 
The GAL model, on the contrary,
yields an error density that has a much thinner left tail,
concentrating more of its probability mass around the mode,
which is not at 0. Figure \ref{fig:pred} shows also the posterior
densities for shape parameter $\gamma$, under a uniform prior in all
cases. For all three quantile regressions, the 95\% posterior credible
interval for $\gamma$ does not include the value of 0, which
corresponds to asymmetric Laplace errors. Median regression is the
only case where 0 is within the effective range of the posterior
distribution for $\gamma$.

\begin{table}[t!]
\vspace{.8em}
    \centering
    \begin{tabular}{l c r r}
        \hline
        & & \multicolumn{2}{c}{Bayesian information criterion}\\\cline{3-4}
        ~Quantile & Model & log-likelihood & BIC\\
        \hline
$~p_0=0.05~~$ & $\text{M}_0$ & $-666$ & $1355$ \\
 & $\text{M}_1$ & $-615$ & $1258$ \\
        \hline
$~p_0=0.25$ & $\text{M}_0$ & $-632$ & $1287$\\
 & $\text{M}_1$ & $-622$ & $1273$ \\
        \hline
$~p_0=0.50$ & $\text{M}_0$ & $-633$ & $1289$ \\
 & $\text{M}_1$ & $-623$ & $1274$ \\
        \hline
$~p_0=0.75$ & $\text{M}_0$ & $-654$ & $1331$\\
 & $\text{M}_1$ & $-620$ & $1268$\\
        \hline
$~p_0=0.95$ & $\text{M}_0$ & $-761$ & $1545$\\
 & $\text{M}_1$ & $-646$ & $1320$\\
        \hline
        \hline
        \multicolumn{4}{l}{}
    \end{tabular}
    \caption{Immunoglobulin-G data. Bayesian information criterion
under the asymmetric Laplace and generalized asymmetric Laplace
models, denoted by $\text{M}_{0}$ and $\text{M}_{1}$, respectively.}
\vspace{-.5em}
    \label{tab:BIC}
\end{table}

For formal model comparison, we compute the
Bayesian information criterion (BIC), the posterior predictive loss 
criterion with quadratic loss, and the generalized posterior predictive 
loss criterion under the check loss.
The Bayesian information criterion favors the new
model at all five quantiles; see Table \ref{tab:BIC}. Under the
posterior predictive loss criterion (Table \ref{tab:PDG}), the two models are comparable
in the case of median regression, with model $\text{M}_{0}$
preferred. In all other cases, model $\text{M}_{1}$ is favored
by both versions of the model comparison criterion.
The improvement in performance over the AL model is particularly
conspicuous at the two extreme percentiles. This is in agreement with
the difference in the posterior predictive error densities
for $p_{0}=0.95$, reported in Figure \ref{fig:pred}.

\begin{table}[th!]
\vspace{.8em}
    \centering
    \begin{tabular}{l c r r r c r r r}
        \hline
        & & \multicolumn{7}{c}{Posterior predictive loss criterion}\\\cline{3-9}
        & & \multicolumn{3}{c}{Quadratic loss} &
        &\multicolumn{3}{c}{Check loss} \\\cline{3-5}\cline{7-9}
        ~Quantile & Model & ~P~~ & ~G~~ & ~D$_\infty$~
        && ~P~~ & ~G~~ & ~D~~\\
        \hline
$~p_0=0.05~~$ & $\text{M}_0$ &$3511$&$1331$&$4841$
    &&$179$&$180$&$359$ \\
 & $\text{M}_1$ &$1298$&$1170$&$2467$
    &&$230$&$102$&$331$\\
        \hline
$~p_0=0.25$ & $\text{M}_0$ &$1820$&$1180$&$3001$
    &&$232$&$123$&$355$ \\
 & $\text{M}_1$ &$1407$&$1144$&$2551$
    &&$236$&$108$&$343$ \\
        \hline
$~p_0=0.50$ & $\text{M}_0$ &$1465$&$1142$&$2607$
    &&$229$&$108$&$338$ \\
 & $\text{M}_1$ &$1626$&$1161$&$2788$
    &&$232$&$114$&$346$ \\
        \hline
$~p_0=0.75$ & $\text{M}_0$ &$2122$&$1227$&$3350$
    &&$201$&$134$&$335$ \\
 & $\text{M}_1$ &$1208$&$1140$&$2348$
    &&$228$&$97$&$325$ \\
        \hline
$~p_0=0.95$ & $\text{M}_0$ &$6522$&$1751$&$8273$
    &&$137$&$259$&$395$ \\
 & $\text{M}_1$ &$1525$&$1165$&$2690$
    &&$208$&$118$&$327$ \\
        \hline
        \hline
        \multicolumn{9}{l}{}
    \end{tabular}
    \caption{Immunoglobulin-G data. Posterior predictive
loss criterion (based on quadratic loss and check loss functions)
under the asymmetric Laplace and generalized asymmetric Laplace
models, denoted by $\text{M}_{0}$ and $\text{M}_{1}$.}
\vspace{-.5em}
    \label{tab:PDG}
\end{table}

\subsection{Boston housing data}
\label{boston}

We apply the lasso regularized quantile regression model to
the realty price data from 
the Boston Standard Metropolitan Statistical Area (SMSA) in 1970
\citep{HarrRubi1978}. The data set contains 506 observations.
We take the log-transformed corrected median value of
owner-occupied housing in USD 1000 (LCMEDV) as the response,
and consider the following
predictors: point longitudes in decimal degrees (LON), point latitudes
in decimal degrees (LAT), per capita crime (CRIM), proportions of
residential land zoned for lots over 25000 square feet per town (ZN),
proportions of non-retail business acres per town (INDUS), a factor
indicating whether tract borders Charles River (CHAS), nitric oxides
concentration (parts per 10 million) per town (NOX), average numbers
of rooms per dwelling (RM), proportions of owner-occupied units built
prior to 1940 (AGE), weighted distances to five Boston employment
centers (DIS), index of accessibility to radial highways per town
(RAD), full-value property-tax rate per USD 10,000 per town (TAX),
pupil-teacher ratios per town (PTRATIO), transformed African American
population proportion (B), and percentage values of lower status
population (LSTAT).

\begin{figure}[t!]
    \centering
    \vspace{-2.5em}
    \includegraphics[width=1.0\linewidth]{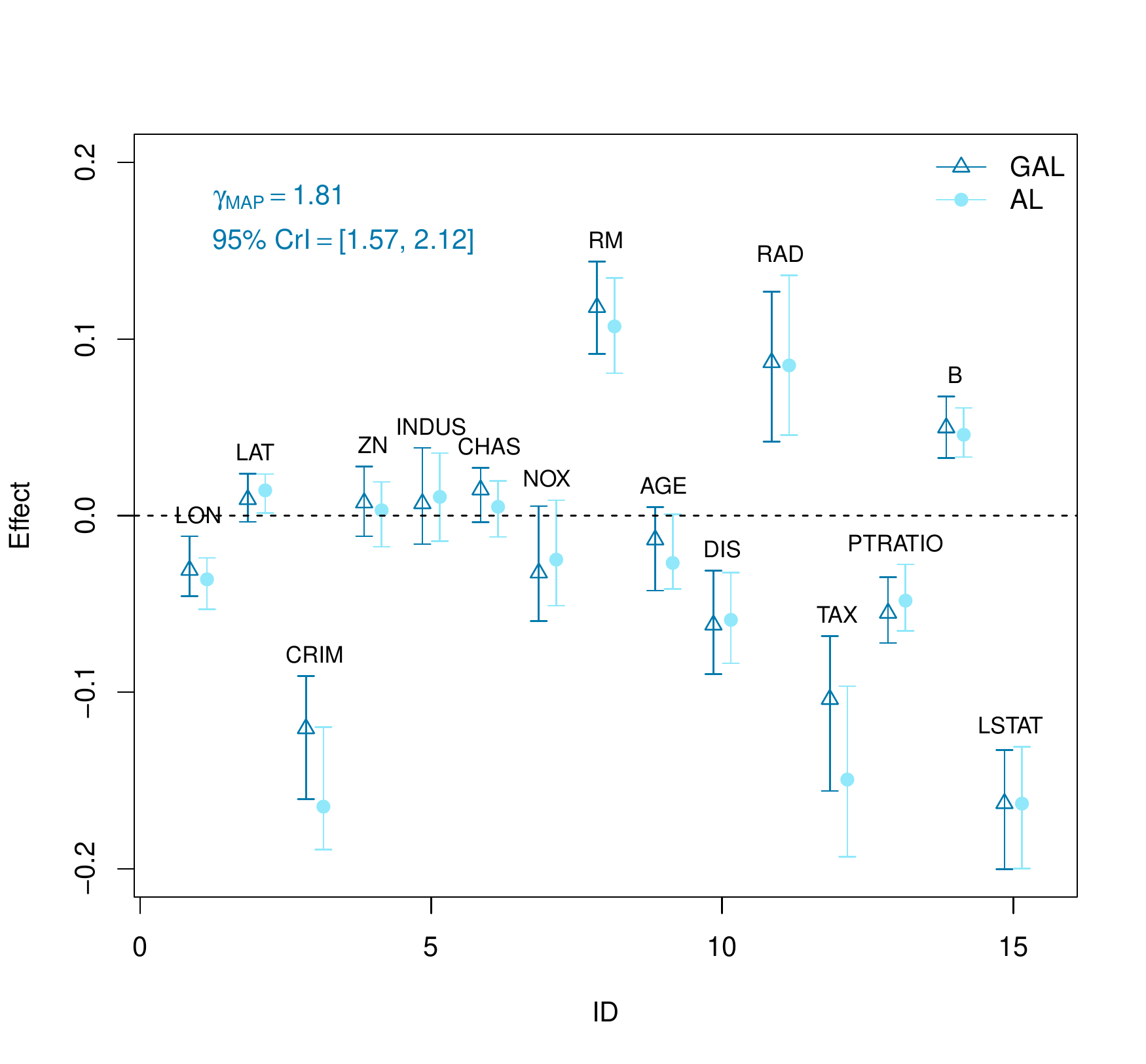}\vspace{-1.3em}
    \captionsetup{margin=2.8em}
    \caption{Boston housing data. Posterior point and 95\% interval
      estimates for the regression coefficients of the 10th quantile
      lasso regularized model under AL and GAL errors.}
    \label{fig:boston10}
\end{figure}

\begin{figure}[t!]
    \centering
    \vspace{-2.5em}
    \includegraphics[width=1.0\linewidth]{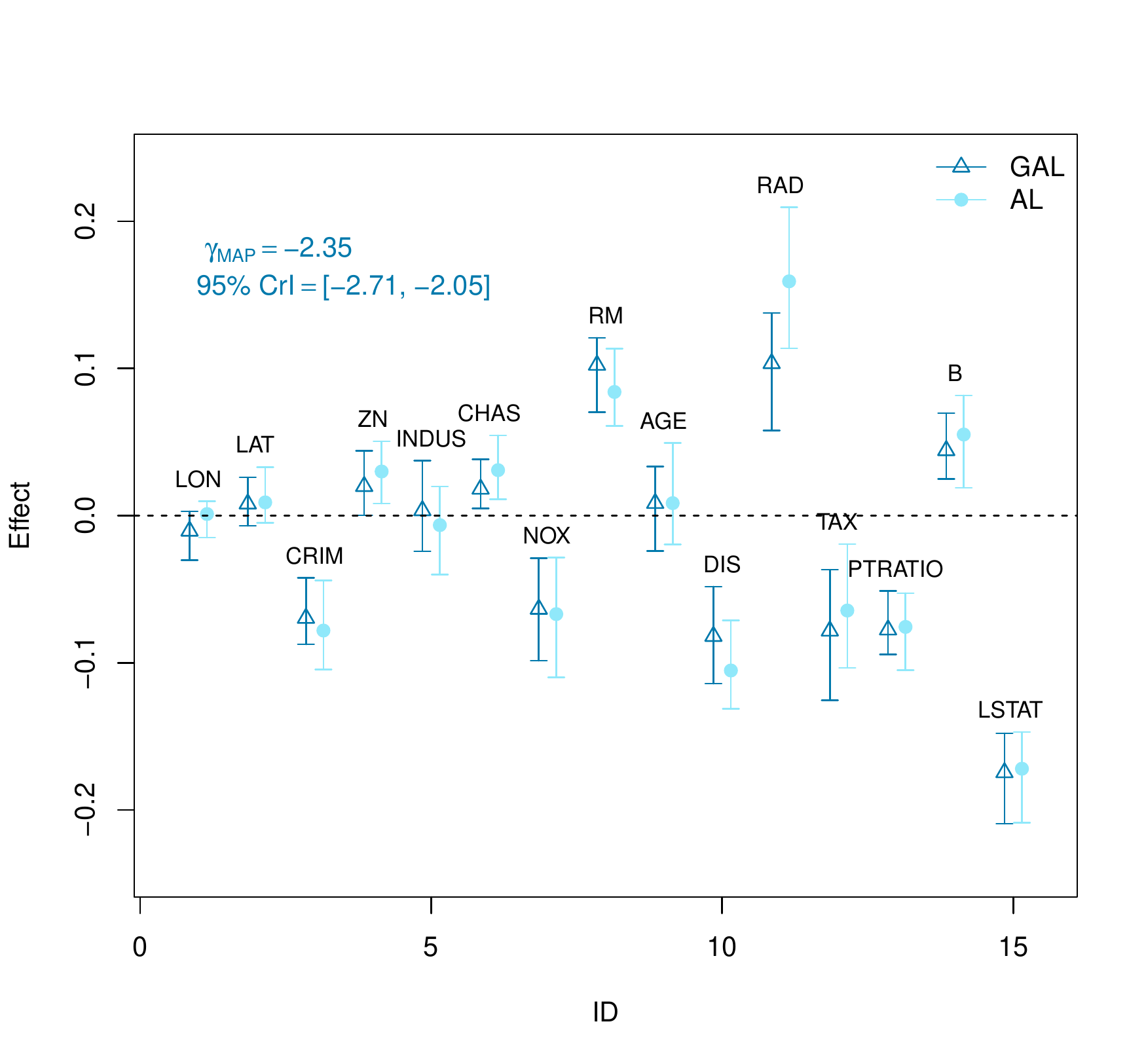}\vspace{-1.3em}
    \captionsetup{margin=2.8em}
    \caption{Boston housing data. Posterior point and 95\% interval
      estimates for the regression coefficients of the 90th quantile
      lasso regularized model under AL and GAL errors.}
    \label{fig:boston90}
\end{figure}

We consider quantiles of $0.1$ and $0.9$ and compare the maximum a
posteriori estimates (MAP) of regression coefficients, along with 95\% credible intervals,
for standardized covariates under the lasso regularized
quantile regression models with AL and GAL errors
(Figure \ref{fig:boston10} and \ref{fig:boston90}). 
For both quantiles, the widths of the 95\% credible intervals for the
regression coefficients are overall comparable between the two 
models, but the posterior point estimates can be quite different. 
For instance, under the
10th quantile regression, the GAL model shrinks the effects of per
capita crime (CRIM) and proporty-tax rate (TAX) to a greater extent
compared to the AL model. Similar patterns can be observed for index
of accessibility to radial highways (RAD) for the 90th quantile.
Moreover, the two models reach different conclusions on
the effect of latitude (LAT) for the 10th percentile. Although the
posterior point estimates suggest a higher housing price as latitude
increases adjusting for all other covariates, the 95\% credible
interval under the GAL model includes 0, whereas the one under
the AL model does not.

Focusing on inference under the GAL error distribution, we note
that, although the model selected some common variables for the two
quantiles, there is also some discrepancy. For instance, each of higher
proportions of residential land zoned for lots over 25000 square feet
per town (ZN) and having tracts bordering Charles river (CHAS) 
increase the price at the 90\% percentile, while higher nitrogen oxide
value (NOX) has a negative influence on the 90\% percentile price. However,
none of these covariates have a significant effect on the realty value
at the 10\% percentile.

Finally, we notice that for both the 10th and 90th quantile
regression, 0 is far away from the
endpoints of the 95\% credible interval for the GAL model
shape parameter $\gamma$. This suggests that asymmetric Laplace errors
are not suitable for this particular application. This is further
supported by the results for the posterior predictive loss criterion
reported in Table \ref{tab:boston_PGD}.

\begin{table}[t!]
\vspace{.8em}
    \centering
    \begin{tabular}{l c r r r c r r r}
        \hline
        & & \multicolumn{7}{c}{Posterior predictive loss criterion}\\\cline{3-9}
        & & \multicolumn{3}{c}{Quadratic loss} &
        &\multicolumn{3}{c}{Check loss} \\\cline{3-5}\cline{7-9}
        ~Quantile & Model & ~P~~ & ~G~~ & ~D$_\infty$~
        && ~P~~ & ~G~~ & ~D~~\\
        \hline
$~p_0=0.10~~$ & $\text{M}_0$ &$46.9$&$26.2$&$73.1$
    &&$28.1$&$22.6$&$50.7$ \\
 & $\text{M}_1$ &$22.8$&$20.1$&$42.9$
    &&$30.4$&$18.5$&$48.9$\\
        \hline
$~p_0=0.90$ & $\text{M}_0$ &$74.8$&$28.8$&$103.6$
    &&$24.1$&$31.5$&$55.7$ \\
 & $\text{M}_1$ &$22.6$&$18.4$&$41.0$
    &&$26.3$&$21.0$&$47.3$ \\
        \hline
        \hline
        \multicolumn{9}{l}{}
    \end{tabular}
    \caption{Boston housing data. Posterior predictive
loss criterion (based on quadratic loss and check loss functions)
under the AL (model $\text{M}_0$) and GAL (model $\text{M}_1$) error distribution.}
\vspace{-.5em}
    \label{tab:boston_PGD}
\end{table}

\subsection{Labor supply data}
\label{labor-supply}

We illustrate the Tobit quantile regression model with the female 
labor supply data from \cite{Mroz1987}, which was taken from the 
University of Michigan Panel Study of Income Dynamics for year 1975.
The data set includes records on the work hours and other relevant 
information of $753$ married white women aged between 30 and 
60 years old. Of the $753$ women, $428$ worked at some time during 1975,
with the corresponding fully observed responses given by the wife's 
work hours (in 100  hours). For the remaining $325$ women, the
observed zero work hours correspond to negative values for the latent
``labor supply'' response. 
We use the quantile regression function considered in \cite{KozuKoba2011},
where an AL-based Tobit quantile regression model was applied to the
same data set. 
The linear predictor includes an intercept, income which is not due to the wife
(\texttt{nwifeinc}), education of the wife in years (\texttt{educ}),
actual labor market experience in years (\texttt{exper}) and its
quadratic term  (\texttt{expersq}), age of the wife (\texttt{age}),
number of children less than 6 years old in household
(\texttt{kidslt6}), and number of children between ages 6 and 18 in
household (\texttt{kidsge6}). We compare the results from
the Bayesian Tobit quantile regression model assuming AL errors 
(model $\text{M}_0$) and GAL errors (model $\text{M}_1$).

%
%
%

Table \ref{tab:PGD_tobit} summarizes the posterior distribution of
$\gamma$ under the GAL model, and presents results from criterion-based
comparison of the two models for $p_0=0.05$,
$0.50$ and $0.95$. Since there is censoring in the data, we use the
revised BIC from \cite{VoliRaft2000}. In all three
cases, the 95\% credible interval for $\gamma$ excludes $0$,
and the GAL-based model is associated with lower BIC values. 
The results support the GAL-based model more emphatically for the extreme
percentiles than for median regression.

\begin{table}[t!]
\vspace{.8em}
    \centering
    \begin{tabular}{lc c rc}
        \hline
        ~Quantile & Model & Mean (95\% CrI) for $\gamma$ & likelihood & BIC 
        \\\hline
$~p_0=0.05$ & $\text{M}_0$ &  
    &$-1975$&$4004$ \\
 & $\text{M}_1$ &$5.22$ ($4.43$, $6.24$)
    &$-1874$&$3809$ \\
        \hline
$~p_0=0.50$ & $\text{M}_0$ &  
    &$-1867$&$3789$ \\
 & $\text{M}_1$ &$0.58$ ($0.39$, $0.81$)
    &$-1845$&$3750$ \\
        \hline
$~p_0=0.95$ & $\text{M}_0$ &  
    &$-1967$&$3989$ \\
 & $\text{M}_1$ &$-4.16$ ($-5.5$, $-3.06$)
    &$-1854$&$3769$ \\
    \hline
    \hline
    \end{tabular}
    \caption{Labor supply data. Posterior mean and 95\% credible
interval for the shape parameter $\gamma$ of the GAL error
distribution, and BIC values under the AL and GAL 
models, denoted by $\text{M}_{0}$ and $\text{M}_{1}$, respectively.}
\vspace{-.5em}
    \label{tab:PGD_tobit}
\end{table}

\begin{figure}[t!]
    \centering
    \vspace{-2.5em}
    \includegraphics[width=1.0\linewidth]{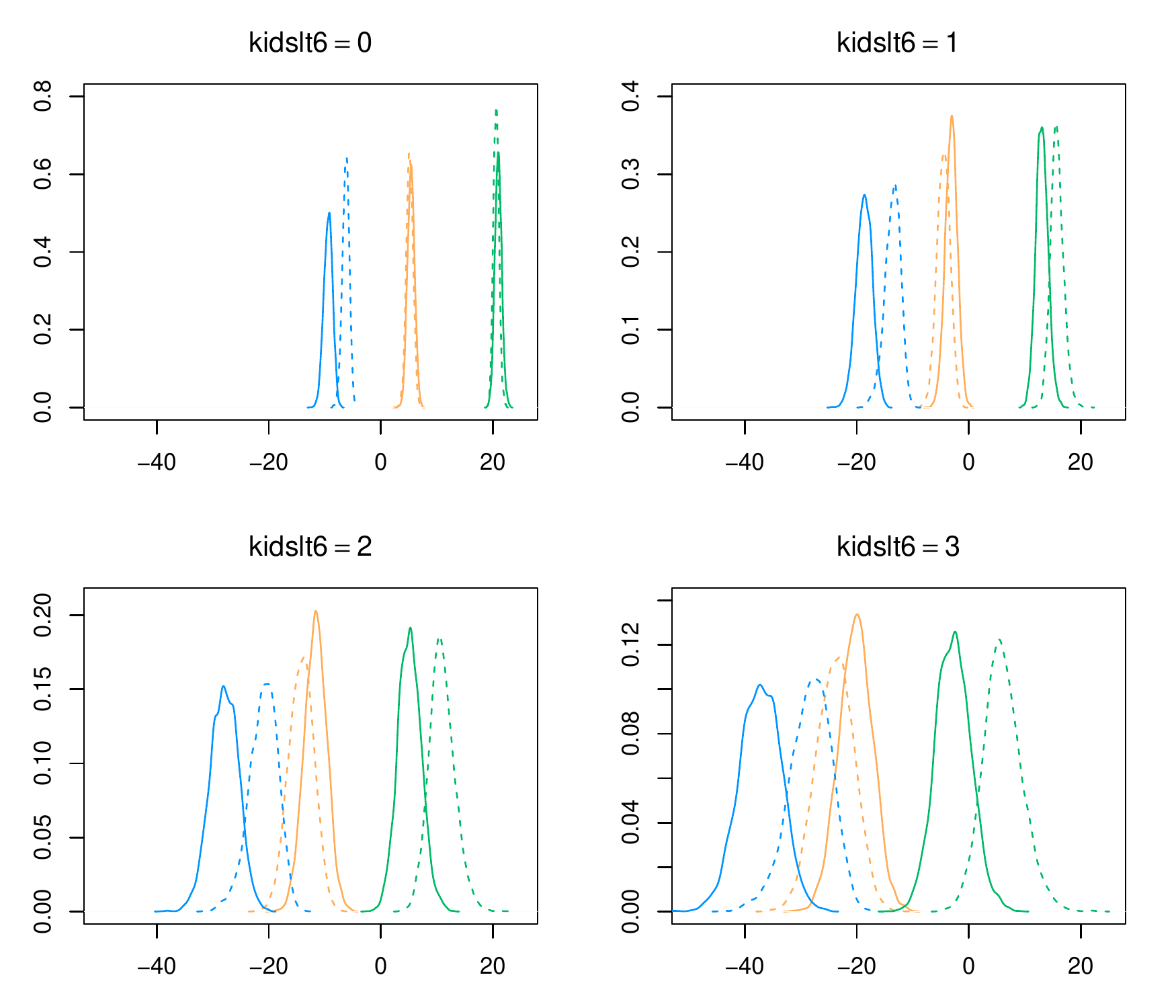}
    \captionsetup{margin=2.8em}
    \caption{Labor supply data. Posterior densities for the 5th (blue), 
50th (orange) and 95th quantile (green) of labor supply (in 100 hours)
for women with 0, 1, 2 or 3 children less than 6 years old. The solid (dashed) 
lines correspond to the posterior densities under the GAL (AL) model.}
    \label{fig:predquant_tobit}
\end{figure}

Figure \ref{fig:predquant_tobit} shows the posterior distributions of 
labor supply quantiles corresponding to $p_0=0.05$, $0.50$ and 
$0.95$ for women with 0, 1, 2 and 3 children less than 6 years old. 
For all other predictors, we use the median values from the data 
as input values to represent an {\it average} wife. 
As the number of young children increases, the AL model estimates 
the 5th quantile and the median of labor supply of an average wife 
to be closer to each other. Under the GAL model, the distance between 
the densities of the 5th quantile and median labor supply also
decreases with increasing number of young children, 
albeit at a lower rate. When estimating the 95th quantile,
the proposed model is more conservative than the AL model about the
labor contribution of an average wife with an increasing number of children
less than 6 years old. When there are 3 children less than 6 years old
in the household, the center of the posterior distribution for the 95th
quantile is below zero under the GAL model, meaning that even at the top 5th percentile of
labor supply, an average wife may still produce negative labor supply
as she takes care of many young family members. More specifically, the
posterior probability of the 95th labor supply quantile being positive 
is 0.19 under the GAL model, as opposed to 0.97 under the AL model.
These results demonstrate that the choice of error distribution in
quantile regression can have an effect on practically important
conclusions for a particular application.

\section{Discussion}

We have developed a Bayesian quantile regression framework with a new error
distribution that has flexible skewness, mode and tail behavior. The
proposed model has better performance compared with the commonly used
asymmetric Laplace distribution, particularly for modeling extreme
quantiles. Owing to the hierarchical structure of the new
distribution, posterior inference and prediction can be readily implemented
via Markov chain Monte Carlo methods.
%

The main motivation for this work was to develop a sufficiently
flexible parametric distribution that can be used as a building block
for different types of quantile regression models.
The extension to quantile regression with ordinal responses is a
possible direction.
Expanding the model to a spatial quantile regression process, along
the lines of \cite{LG2012}, is another direction. Finally, current work is
exploring a composite quantile regression modeling framework,
built from structured mixtures of generalized asymmetric Laplace
distributions, to combine information from multiple quantiles of the
response distribution in inference for variable selection.

\section*{Acknowledgements}
This research was supported in part by the National Science Foundation
under award SES-1631963.

\bibliographystyle{asa}
\bibliography{bibliography}

\end{document}